\documentclass[12pt,epsf]{article}
\usepackage{axodraw}
\usepackage{graphicx}
\usepackage{enumerate} 
\def\lsim{\mathrel{\vcenter{\hbox{$<$}\nointerlineskip\hbox{$\sim$}}}}
\newcommand{\be}{\begin{equation}}
\newcommand{\ee}{\end{equation}}
\newcommand{\ba}{\begin{eqnarray}}
\newcommand{\ea}{\end{eqnarray}}

\def\21{$SU(2) \otimes U(1) $}

\def\lsim{\raise0.3ex\hbox{$\;<$\kern-0.75em\raise-1.1ex\hbox{$\sim\;$}}}
\def\gsim{\raise0.3ex\hbox{$\;>$\kern-0.75em\raise-1.1ex\hbox{$\sim\;$}}} 

\newcommand{\mx}{\left[\begin{array}}
\newcommand{\finmx}{\end{array}\right]} 
\newcommand{\mxp}{\left(\begin{array}} 
\newcommand{\finmxp}{\end{array}\right)} 
\def\beq{\begin{equation}}
\def\eeq{\end{equation}}
\def\bea{\begin{eqnarray}}
\def\eea{\end{eqnarray}}

\def\mathbf#1{\hbox{\bf #1}}
\def\textrm#1{\hbox{#1}}

\def\lsim{\raise0.3ex\hbox{$\;<$\kern-0.75em\raise-1.1ex\hbox{$\sim\;$}}}
\def\gsim{\raise0.3ex\hbox{$\;>$\kern-0.75em\raise-1.1ex\hbox{$\sim\;$}}}
\newcommand {\ignore}[1]{}

\begin{document}
\vspace*{-1in}
\renewcommand{\thefootnote}{\fnsymbol{footnote}}
\begin{flushright}
\texttt{
} 
\end{flushright}
\vskip 5pt
\begin{center}
{\Large{\bf The SUSY seesaw model and lepton-flavor violation at 
a future electron-positron linear collider}}
\vskip 25pt

{\sf 
F. Deppisch$^{1,2}$\footnote[1]{E-mail: deppisch@physik.uni-wuerzburg.de}, 
H. P\"as$^1$\footnote[2]{E-mail: paes@physik.uni-wuerzburg.de}, 
A. Redelbach$^1$\footnote[3]{E-mail: asredelb@physik.uni-wuerzburg.de}, 
R. R\"uckl$^1$\footnote[4]{E-mail: rueckl@physik.uni-wuerzburg.de},
Y. Shimizu$^3$\footnote[5]{E-mail: shimizu@eken.phys.nagoya-u.ac.jp}}
\vskip 10pt
{\it \small $^1$ Institut f\"ur Theoretische Physik und Astrophysik\\
Universit\"at W\"urzburg\\ D-97074 W\"urzburg, Germany}\\
{\it \small 
$^2$ Institut de F\'{\i}sica Corpuscular - C.S.I.C., 
Universitat de Val{\`e}ncia \\
Edifici Instituts d'Investigaci{\'o} - Apartat de Correus 22085 - 46071 
Val{\`e}ncia, Spain}\\
{\it \small $^3$ Department of Physics \\ Nagoya University\\ Nagoya, 464-8602, 
Japan}\\

\vskip 20pt

{\bf Abstract}
\end{center}

\begin{quotation}
{\small 
We study lepton-flavor violating slepton production and decay 
at a future $e^+e^-$ linear collider in context with the 
seesaw mechanism in mSUGRA post-LEP benchmark scenarios. 
The present knowledge in the neutrino sector
as well as improved future measurements 
are taken into account. 
We calculate the signal cross-sections
$\sigma(e^{\pm}e^-\rightarrow l_{\beta}^{\pm} l_{\alpha}^- 
\tilde{\chi}_b^0 \tilde{\chi}_a^0)$; $l_{\delta}=e, \mu, \tau$;
$\alpha \neq \beta$
and estimate the main background processes. Furthermore,
we investigate the correlations of these signals with the corresponding 
lepton-flavor violating rare decays $l_{\alpha} \rightarrow l_{\beta} \gamma$. 
It is shown that these correlations are relatively weakly affected by uncertainties 
in the neutrino data, but very sensitive to the model parameters. Hence, they
are particularly suited for probing the origin of lepton-flavor violation.
}
\end{quotation}

\vskip 20pt  

\setcounter{footnote}{0}
\renewcommand{\thefootnote}{\arabic{footnote}}

\newpage
\section{Introduction}
One of the main virtues of experiments at an $e^+e^-$ linear 
collider (LC) is the clean environment allowing studies of
the production and decay of new particles with low background.
This not only enables precision measurements of particle properties,
but also searches for very rare processes and small effects. 
An important example of this kind is lepton-flavor 
violation (LFV) as suggested by the experimental evidence 
for neutrino oscillations \cite{sno,sk} 
and expected particularly in SUSY models.  
Phenomenological investigations have indicated how 
tests of LFV at a high-energy LC could nicely complement searches for 
lepton-flavor violating rare decays such as 
$\mu \rightarrow e \gamma$.
Previous work \cite{Arkani-Hamed:1996au} 
has mainly focussed on slepton-pair production assuming
two-generation slepton mixing. 
LFV in chargino-pair production was analyzed in \cite{Guchait:2001us} 
together with a detailed background analysis. Recently, it was shown 
in a rather model-independent analysis \cite{Porod:2002zy}
by scanning over all possible soft SUSY-breaking terms
consistent with the existing low-energy bounds that 
LFV in the above processes may be quite sizable, leading to final states 
with a pair of charged leptons of unequal flavor, missing energy,
and possibly additional leptons and jets.

In supersymmetric theories with heavy right-handed Majorana neutrinos, 
the seesaw mechanism \cite{seesaw}
can give rise to light neutrino masses at or below 
the sub-eV scale. Moreover, the massive neutrinos affect the 
renormalization group running of the slepton masses, which leads 
to mixing of different slepton flavors.
In the present article we investigate the implications of recent
neutrino measurements on this mixing, extending our 
previous analysis of the radiative decays 
$l_{\alpha} \rightarrow l_{\beta} \gamma$ 
in the SUSY seesaw model \cite{Deppisch:2002vz} 
to the lepton-flavor violating processes 
$e^{\pm}e^-\rightarrow l_{\beta}^{\pm} 
l_{\alpha}^- \tilde{\chi}_b^0 \tilde{\chi}_a^0$ 
involving slepton-pair production and subsequent decay. 
We again use the mSUGRA benchmark scenarios proposed in 
\cite{Battaglia:2001zp} for LC studies, concentrating on those
which predict charged sleptons that are light enough to be pair-produced 
at the center-of-mass energy $\sqrt{s}=500$~GeV.
Furthermore, we examine the most important background processes 
and indicate the most promising lepton-flavor violating channels.
The fact that in the benchmark models considered LFV occurs only in the 
left-handed slepton sector makes the separation of signal and background
more difficult than in models with LFV in the right-handed sector. 
Therefore, our choice of models provides good study cases for developing 
search strategies.
Finally, we work out the correlations between LFV  
in the high-energy $e^\pm e^-$ collisions and the radiative lepton decays,
and show that these correlations are relatively weakly affected by the 
uncertainties in the neutrino data, but very sensitive to the mSUGRA parameters.
Consequently, they could play an important role in probing the class of
models of LFV studied here.

The paper is organized as follows. In section~2 we briefly outline
the SUSY seesaw mechanism and discuss the slepton and neutralino mass 
matrices. 
Section~3 summarizes our analytic results on the helicity amplitudes 
for slepton-pair production, 
$e^\pm e^- \to \tilde{l}_j^{\pm}\tilde{l}^-_i$, and slepton decay,
$\tilde{l}_i^\pm\to l_\alpha^\pm\tilde{\chi}_a^0$, and provides the 
cross-section formula for the complete $2 \rightarrow 4$ processes 
$e^{\pm}e^- \rightarrow l_{\beta}^{\pm}
l_{\alpha}^- \tilde{\chi}_b^0 \tilde{\chi}_a^0$.
The fundamental parameters of the mSUGRA benchmark scenarios 
as well as the neutrino data used in our numerical analysis are specified 
in section~4. In this section we also present our predictions
on the signal cross-sections, while the estimates of the background and 
the effects of energy and angular cuts are given in section~5. 
The prospects for LC searches anticipated on the basis of our analysis are 
summarized in section~6.

\section{Lepton-flavor violation in the charged slepton sector}

\subsection{Supersymmetric seesaw mechanism}

In supersymmetric models the existence of right-handed
neutrino singlet fields with Majorana mass terms and Yukawa interactions
gives rise to the well-known seesaw mechanism, provided the Majorana mass 
scale $M_R$ is much higher than the scale \(v=174\)~GeV of electroweak 
symmetry breaking. At energies below $M_R$ the relevant term in the  
effective superpotential is given by \cite{Casas:2001sr}
\beq
W^{eff}_{\nu} = \frac{1}{2}(Y_\nu L \cdot  H_2 )^T M^{-1}(Y_\nu L \cdot
H_2),
\eeq
where \(L\) denotes the left-handed lepton doublets, $H_{2}$ 
the Higgs doublet with hypercharge $+\frac{1}{2}$, and \(Y_\nu\) the
matrix of neutrino Yukawa couplings.
After electroweak symmetry breaking, the Yukawa couplings 
generate the Dirac mass matrix \(m_D=Y_\nu \langle H_2^0 \rangle\), 
\(\langle H_2^0 \rangle = v\sin\beta\) being the $H_{2}$ vacuum
expectation value with 
\(\tan\beta = \frac{\langle H_2^0\rangle}{\langle H_1^0\rangle}\).
This in turn leads to the Majorana mass matrix
\beq\label{eqn:SeeSawFormula}
M_\nu = m_D^T M^{-1} m_D = Y_\nu^T M^{-1} Y_\nu (v \sin\beta )^2
\eeq
for the light neutrinos which is diagonalized by the 
unitary MNS matrix \(U\):
\beq\label{eqn:NeutrinoDiag}
U^T M_\nu U = \textrm{diag}(m_1, m_2, m_3).
\eeq
The matrix $U$ and the mass eigenvalues $m_i$ are constrained by neutrino 
data as discussed, e.g., in \cite{Deppisch:2002vz}.

\subsection{Renormalization group evolution of the slepton masses}

The heavy neutrino mass eigenstates contribute to the renormalization group 
running of the slepton mass matrices thereby inducing flavor non-diagonal terms which 
are responsible for the lepton-flavor violating processes described later.
At the unification scale $M_{X}$ we assume the mSUGRA universality conditions
\begin{equation}
m^{2}_{\tilde{l}_{L}}=m_{0}^{2}\mathbf{1},\qquad m^{2}_{\tilde{l}_{R}}=m_{0}^{2}\mathbf{1},
\qquad A =A_{0}Y_{l},
\end{equation}
where $m_{0}$ is the common scalar mass and $A_{0}$ the common trilinear coupling.
At lower scales of order of the SUSY threshold, the mass squared matrix of the 
charged sleptons has the form
\begin{equation}\label{ch_slepton_mass_mat}
  m_{\tilde l}^2=\left(
    \begin{array}{cc}
        m^2_{\tilde{l}_{L}}    & m^{2\;\dagger}_{\tilde{l}_{LR}} \\
        m^{2}_{\tilde{l}_{LR}} & m^{2}_{\tilde{l}_{R}}
    \end{array}
      \right),
\end{equation}
where  \(m^2_{\tilde{l}_{L}}\), \(m^{2}_{\tilde{l}_{R}}\) and \(m^{2}_{\tilde{l}_{LR}}\) 
are \(3\times3\) matrices in flavor space, 
\(m^2_{\tilde{l}_{L}}\) and \(m^{2}_{\tilde{l}_{R}}\) being hermitian.
The respective matrix elements are given by
\begin{eqnarray}
  (m^2_{\tilde{l}_L})_{ij}     &=& (m_{L}^2)_{ij} + \delta_{ij}\left(m_{l_i}^2 + 
m_Z^2 \cos2\beta\left(-\frac{1}{2}+\sin^2\theta_W \right)\right) 
\label{slepcorr_1}\\
  (m^2_{\tilde{l}_{R}})_{ij}     &=& (m_{R}^2)_{ij} + \delta_{ij}(m_{l_i}^2 - 
m_Z^2 \cos2\beta\sin^2\theta_W) \label{slepcorr_2}\\
 (m^{2}_{\tilde{l}_{LR}})_{ij} &=& A_{ij}v\cos\beta-\delta_{ij}m_{l_i}\mu\tan\beta,
\label{slepcorr_3}
\end{eqnarray}
\(\theta_W\) being the weak-mixing 
angle, and \(\mu\) the SUSY Higgs-mixing parameter. 
The contributions to
the first terms on the r.h.s. of (\ref{slepcorr_1}) - (\ref{slepcorr_3})  
are indicated below:
\begin{eqnarray}
m_{L}^2&=&m_0^2\mathbf{1} + (\delta m_{L}^2)_{\textrm{\tiny MSSM}} + \delta m_{L}^2 
\label{left_handed_SSB} \\
m_{R}^2&=&m_0^2\mathbf{1} + (\delta m_{R}^2)_{\textrm{\tiny MSSM}} + \delta m_{R}^2 
\label{right_handed_SSB}\\
A&=&A_0 Y_l+\delta A_{\textrm{\tiny MSSM}}+\delta A \label{A_SSB}.
\end{eqnarray}
Here, $(\delta m^{2}_{L,R})_{\textrm{\tiny MSSM}}$ and $\delta A_{\textrm{\tiny MSSM}}$ 
denote the usual MSSM renormalization group corrections \cite{deBoer:1994dg}
which are flavor-diagonal.
In addition, the right-handed neutrinos radiatively 
induce the flavor off-diagonal terms $\delta m_{L,R}$ and $\delta A$.
In leading logarithmic approximation
\footnote{The exact one-loop renormalization group equations may give somewhat different
results as pointed out in \cite{Petcov:2003zb} and studied in more detail for  
lepton-flavor violating processes of interest in a forthcoming 
article \cite{Deppisch:2004wz}. For consistency with the results on radiative decays, 
derived in \cite{Deppisch:2002vz} and used in the present analysis,
we restrict ourselves here to the approximation 
(\ref{eq:rnrges})-(\ref{eq:rnrges123}).},
one gets \cite{Hisano:1999fj}
\begin{eqnarray}\label{eq:rnrges}
  \delta m_{L}^2 &=& -\frac{1}{8 \pi^2}(3m_0^2+A_0^2)(Y_\nu^\dag L Y_\nu) 
\\
  \delta m_{R}^2 &=& 0  \\
  \delta A &=& -\frac{3 A_0}{16\pi^2}(Y_l Y_\nu^\dag L Y_\nu)
\label{eq:rnrges123}
\end{eqnarray}
with
\beq
L_{ij}=\ln\left(\frac{M_X}{M_{i}}\right)\delta_{ij},
\eeq
$M_i,~i=1,2,3$ being the eigenvalues of the Majorana mass matrix $M$ which 
may be chosen diagonal.
Using (\ref{eqn:SeeSawFormula}) and (\ref{eqn:NeutrinoDiag}), the Yukawa matrix $Y_\nu$
can be parametrized as follows \cite{Casas:2001sr}:
\beq
Y_\nu=\frac{1}{v\sin\beta}\textrm{diag}\left(\sqrt{M_1}, \sqrt{M_2}, \sqrt{M_3}\right) R \; 
\textrm{diag}\left(\sqrt{m_1}, \sqrt{m_2}, \sqrt{m_3}\right)U^\dagger, 
\label{eq:yukawa}
\eeq
where $R$ is an undetermined complex orthogonal matrix.
For substitution in (\ref{eq:rnrges}) and (\ref{eq:rnrges123}) 
the Yukawa couplings have to be evolved from the low-energy scale 
taken to be $M_Z$ to the Majorana scale $M_R$ and further to the GUT scale $M_X$.

Following \cite{Deppisch:2002vz} and many previous studies,
we assume degenerate Majorana masses $M_{1,2,3}=M_R$ and take $R$ to be real.
In this case, $R$
drops out and the product $ Y_\nu^\dagger L Y_\nu$
is simply given by
\begin{eqnarray}\label{eqn:yy}
 Y_\nu^\dagger L Y_\nu = Y_\nu^\dagger Y_\nu \ln\left(\frac{M_X}{M_{R}}\right) 
 = \frac{M_R}{v^2\sin^2\beta}U \cdot
\textrm{diag}(m_1, m_2, m_3) \cdot U^\dagger\ln\left(\frac{M_X}{M_{R}}\right).
\label{yukawprod}
\end{eqnarray}
A more general investigation allowing for non-degenerate Majorana masses 
is in progress and will be reported elsewhere \cite{Deppisch:2004vz}.
Here, we stick to (\ref{yukawprod}) since we want to
investigate the correlations between the 
high-energy processes, which are the main subject of the present paper,
and the low-energy rare decays studied in \cite{Deppisch:2002vz}. There,
one can find some preliminary qualitative remarks on the effects of a 
complex matrix $R$.
 
From (\ref{ch_slepton_mass_mat}) the mass
eigenvalues $m_{\tilde l_i},~i=1,...,6$
of the sleptons 
are obtained via diagonalization by a \(6\times6\) unitary matrix \(U_{\tilde l}\):
\begin{equation}
  U_{\tilde l}^\dagger m_{\tilde l}^2 U_{\tilde l}=\textrm{diag}(m_{\tilde l_1}^2,...,
m_{\tilde l_i}^2,...,m_{\tilde l_6}^2).
\end{equation}
For definiteness, the eigenvalues are ordered such that the masses increase 
from $\tilde l_1$ to $\tilde l_6$. In other words, in the absense of LFV
$\tilde l_1= \tilde \tau_1$, $\tilde l_2= \tilde \mu_1$, ..., $\tilde l_6
= \tilde \tau_2$.
The corresponding mass
eigenstates are then expressed in terms of the gauge eigenstates by
\begin{equation}
  \tilde l_i=(U_{\tilde l}^*)_{\alpha i}\tilde l_{L\alpha} + (U_{\tilde l}^*)_{(\alpha+3)i} 
\tilde l_{R\alpha}, \quad\quad i=1,...,6; \;\alpha=e,\mu,\tau.
\label{slepstates}
\end{equation}

\subsection{Neutralino mass matrix}
In slepton production and decay neutralinos play an important role. Therefore,
we want to clarify our notation for later use. The physical neutralinos are mixtures of
gauginos and higgsinos.
In the gauge-eigenstate basis $\psi^0=(\tilde{B},\tilde{W}^0,\tilde{H}^0_1,\tilde{H}^0_2)^T$, 
the neutralino mass term reads \cite{Haber:1984rc} 
\begin{equation}
\mathcal{L}_{\psi}=-\frac{1}{2}(\psi^0)^T M_{\psi} \psi^0 + h.c.,
\end{equation}
with
\begin{equation}
M_{\psi}=\left(
\begin{tabular}{cccc}
$M_1$ & 0       & $-m_Z s_W c_\beta$ & $m_Zs_Ws_\beta$ \\
0   & $M_2$     & $m_Zc_W c_\beta$   & $-m_Z c_W s_\beta$ \\ 
$-m_Zs_Wc_\beta$ & $m_Zc_W c_\beta$ & 0 & $-\mu$ \\ 
$m_Zs_Ws_\beta$  & $-m_Zc_Ws_\beta$ & $-\mu$ & 0 
\end{tabular}
\right),
\end{equation}
$M_1$ being the $U(1)$ and $M_2$ the $SU(2)$ gaugino mass. The abbreviations 
$s_\phi = \sin\phi$ and $c_\phi = \cos\phi$ introduced above will be used 
throughout the paper.

The neutralino mass eigenstates are given by
\begin{equation}
\chi^0_a=N_{ab}\psi^0_b, \qquad a,b=1,\ldots,4,
\label{neutstates}
\end{equation}
where $N$ is the unitary matrix diagonalizing $M_{\psi}$:
\begin{equation}
N^* M_{\psi} N^{-1}
=\textrm{diag}\left(m_{\tilde{\chi}^0_1},m_{\tilde{\chi}^0_2},m_{\tilde{\chi}^0_3},
m_{\tilde{\chi}^0_4}\right),
\end{equation}
and $m_{\tilde{\chi}^0_a}$, $a=1,...,4$ are the mass eigenvalues. Finally,
the Majorana spinors for the neutralinos are composed of the Weyl spinors (\ref{neutstates})
as follows:
\begin{equation}
\tilde{\chi}^0_a=\left(
                        \begin{tabular}{c} 
                        $\chi_a^0$\\
                        $\bar{\chi}_a^0$
                        \end{tabular}
                 \right).
\end{equation}

\section{Amplitudes and cross-sections}
The flavor mixing in the slepton sector induced by the heavy neutrinos 
as outlined in
section~2.2 gives rise to the lepton-flavor violating processes 
$e^\pm e^- \to \tilde{l}_j^{\pm}\tilde{l}^-_i\to 
l_\beta^{\pm}l^-_\alpha\tilde{\chi}^0_b\tilde{\chi}^0_a$, $i,j=1,...,6$.
More specifically, in these processes LFV is caused by
the slepton mixing matrix \(U_{\tilde l}\) in (\ref{slepstates}), 
which enters both the slepton production and the decay vertices. 
As a consequence, factorization
in production cross-section times branching ratios is not appropriate. 
One rather has to coherently sum over the intermediate slepton states.
In the following, we summarize our analytical results on the amplitudes and 
cross-sections
for the above processes. The detailed calculations have been 
carried out in \cite{Deppisch:Dip} 
and \cite{Redelbach:Dip}. 

\subsection{The lepton-flavor violating vertices}
The lowest-order Feynman diagrams for the processes under consideration 
are shown in Fig.~\ref{e+e-_diags} and  Fig.~\ref{e-e-_diags}
with the particle four-momenta being defined in brackets. 
While the $\gamma \tilde{l}_{i}^+\tilde{l}_{i}^-$ vertex appearing 
only in Fig.~\ref{e+e-_diags} is flavor-diagonal, LFV occurs in both 
the \(Z\tilde{l}_j^+\tilde{l}_i^-\) and
\(l_\alpha^-\tilde{l}_i^-\tilde{\chi}_a^0\) vertices described by 
\begin{equation}
-i\frac{e}{c_Ws_W}z_{ij}(p_3-p_4)^\mu 
\end{equation}
with
\begin{equation}
z_{ij}= -\frac{1}{2}\sum_{\alpha=1}^3 (U_{\tilde l})_{i\alpha} 
(U^*_{\tilde l})_{j\alpha}+s^2_W \delta_{ij}
\end{equation}
and 
\begin{equation}
-\sqrt{2}ie\left(A_{i\alpha,a} P_L +B_{i\alpha,a}P_R\right)
\label{chill}
\end{equation}
with
\begin{eqnarray}
A_{i\alpha,a} &=&\frac{m_\alpha N_{a3}^*}{2m_W s_W c_\beta} 
                  (U_{\tilde l})_{i(\alpha+3)} 
                  - \frac{N_{a1}^* t_W +N_{a2}^*}{2 s_W} (U_{\tilde l})_{i\alpha}\\
B_{i\alpha,a} &=& \frac{N_{a1}}{c_W} (U_{\tilde l})_{i(\alpha+3)} 
                  + \frac{m_\alpha N_{a3}}{2m_W s_W c_\beta} (U_{\tilde l})_{i\alpha},
\end{eqnarray}
respectively. Here, $P_{L,R}=\frac{1}{2}(1\mp \gamma_5)$ are the left-right projectors,
while the matrix elements $N_{ab}$ are defined in (\ref{neutstates}), and $U_{i\alpha}$ 
in (\ref{slepstates}). The \(l_\beta^+\tilde{l}_j^+\tilde{\chi}_b^0\) vertex is obtained 
from (\ref{chill}) after replacing 
$A_{i\alpha,a}$ by $B^*_{j\beta,b}$ and $B_{i\alpha,a}$ by $A^*_{j\beta,b}$.

\subsection{Helicity amplitudes for slepton production and decay}

Although we will not consider $e^{\pm}$-beam polarization in the 
subsequent numerical analysis, we nevertheless present here the 
helicity amplitudes for future studies of
polarization effects. In comparison to the slepton and neutralino masses,
and to the cms energy of a LC the masses of the initial and final state 
leptons can be neglected in the kinematics.
In terms of the Mandelstam variables \(s=(p_1+p_2)^2\), \(t=(p_1-p_3)^2\), 
\(u=(p_1-p_4)^2\) the helicity amplitudes \(M_{ij}(h_{e^-},h_{e^+})\) (\(h=\pm 1\)) 
for the production process $e^+e^-\to \tilde{l}_j^+\tilde{l}^-_i$
are then given by
\begin{eqnarray}
M_{ij}(+,+)&=&-2ie^2\sqrt{s}\sum_{c=1}^4m_{\tilde{\chi}^0_c}
              \frac{B_{i1,c}A^*_{j1,c}}{t-m_{\tilde{\chi}^0_c}^2} \nonumber\\
M_{ij}(-,-)&=&-2ie^2\sqrt{s}\sum_{c=1}^4m_{\tilde{\chi}^0_c}
              \frac{A_{i1,c}B^*_{j1,c}}{t-m_{\tilde{\chi}^0_c}^2} \nonumber\\
M_{ij}(+,-)&=&-2ie^2\sqrt{tu-p_3^2p_4^2}\!\left(\frac{\delta_{ij}}{s}
              +\frac{1}{c^2_W} \frac{z_{ij}}{s-m_Z^2} 
              +\sum_{c=1}^4\frac{B_{i1,c}B^*_{j1,c}}{t-m_{\tilde{\chi}^0_c}^2}\right) 
              \nonumber\\
M_{ij}(-,+)&=&+2ie^2\sqrt{tu-p_3^2p_4^2}\!\left(\frac{\delta_{ij}}{s}
              +\frac{s^2_W-\frac{1}{2}}{s^2_W c^2_W}\frac{z_{ij}}{s-m_Z^2} 
              +\sum_{c=1}^4\frac{A_{i1,c}A^*_{j1,c}}{t-m_{\tilde{\chi}^0_c}^2}\right).
\end{eqnarray}

Analogously, the helicity amplitudes \(M_{ij}(h_{e^-},h_{e^-})\) for 
$e^-e^-\to \tilde{l}_i^-\tilde{l}^-_j$ read:

\begin{eqnarray}
M_{ij}(+,+)&=& 2ie^2\sqrt{s}\sum_{c=1}^4m_{\tilde{\chi}^0_c}
               \frac{B_{i1,c}B_{j1,c}}{t-m_{\tilde{\chi}^0_c}^2}
               +(t\to u, i\leftrightarrow j)\nonumber\\
M_{ij}(-,-)&=& 2ie^2\sqrt{s}\sum_{c=1}^4m_{\tilde{\chi}^0_c}
               \frac{A_{i1,c}A_{j1,c}}{t-m_{\tilde{\chi}^0_c}^2} 
               +(t\to u, i\leftrightarrow j)\nonumber\\
M_{ij}(+,-)&=& 2ie^2\sqrt{tu-p_3^2p_4^2}\sum_{c=1}^4
               \frac{B_{i1,c}A_{j1,c}}{t-m_{\tilde{\chi}^0_c}^2}
               -(t\to u, i\leftrightarrow j)\nonumber\\
M_{ij}(-,+)&=& -2ie^2\sqrt{tu-p_3^2p_4^2}\sum_{c=1}^4
               \frac{A_{i1,c}B_{j1,c}}{t-m_{\tilde{\chi}^0_c}^2}
               -(t\to u, i\leftrightarrow j).
\end{eqnarray}


Furthermore, for the decay \(\tilde{l}_i^-\to l_\alpha^-\tilde{\chi}_a^0\) 
the helicity amplitudes \(M_i^-(h_{l^-_\alpha})\) summed over the
helicity of the neutralino are given by
\begin{eqnarray}
M_i^-(+) &=& 2ie B^*_{i\alpha,a} \sqrt{p_\alpha\cdot p_a} \nonumber\\
M_i^-(-) &=& 2ie A^*_{i\alpha,a} \sqrt{p_\alpha\cdot p_a}.
\label{hel_decays}
\end{eqnarray}
From these one can obtain the helicity amplitudes \(M_j^+(h_{l^+_\beta})\) for 
\(\tilde{l}_j^+ \to l_\beta^+\tilde{\chi}_b^0\) by the substitutions 
\(B_{i\alpha,a} \to A^*_{j\beta,b}\) 
and \(A_{i\alpha,a}\to B^*_{j\beta,b}\) \cite{Haber:1984rc}.

\subsection{Cross-sections}

Having the amplitudes at hand it is straightforward to derive the
cross-sections  for the complete $2\to4$ processes 
$e^\pm e^- \to \tilde{l}_j^{\pm}\tilde{l}^-_i\to 
l_\beta^{\pm}l^-_\alpha\tilde{\chi}^0_b\tilde{\chi}^0_a$.  
For the square of the amplitudes
summed over all possible intermediate slepton states
one finds
\begin{eqnarray}
|M|^2\!\!&=&\!\!\sum_{ijkl}(M_{ij} M_{kl}^{*})(M_i^- M_k^{-*})(M_j^{\pm} M_l^{\pm*})
              C_{ik}\frac{\pi}{2\overline{m\Gamma}_{ik}}
              C_{jl}\frac{\pi}{2\overline{m\Gamma}_{jl}} \nonumber\\
      &\quad& \times \left(\delta(p_3^2-m_{\tilde l_i}^2)
              +\delta(p_3^2-m_{\tilde l_k}^2)\right) 
              \left(\delta(p_4^2-m_{\tilde l_j}^2)
              +\delta(p_4^2-m_{\tilde l_l}^2)\right)\label{full_M_squared},
\label{xsection}
\end{eqnarray}
with
\begin{eqnarray}
C_{ik}&=&\frac{1}{1+i\frac{\Delta \tilde m_{ik}^2}{2\overline{m\Gamma}_{ik}}},
\nonumber\\
\overline{m\Gamma}_{ik}&=&\frac{1}{2}(m_{\tilde l_i}\Gamma_{\tilde l_i}
                          +m_{\tilde l_k}\Gamma_{\tilde l_k}), 
\qquad  \Delta \tilde m_{ik}^2=m_{\tilde l_i}^2-m_
                               {\tilde l_k}^2. \label{flavor_correlations}
\end{eqnarray}
Here, we have used the narrow width approximation for the slepton propagators, 
which is well justified since the slepton widths $\Gamma_{\tilde l_i}$ are of 
order GeV or less, that is much smaller than the
slepton masses of order 100~GeV. For the product of two slepton propagators,
this approximation yields 
(see, e.g., \cite{Arkani-Hamed:1996au})
\be
\left(\frac{1}{p^2-m_{\tilde l_i}^2+im_{\tilde l_i}\Gamma_{\tilde l_i}}\right)
\left(\frac{1}{p^2-m_{\tilde l_k}^2+im_{\tilde l_k}\Gamma_{\tilde l_k}}\right)^* 
\approx C_{ik} \frac{\pi}{2 \overline{m\Gamma}_{ik}}
\left(\delta(p^2-m_{\tilde l_i}^2)+\delta(p^2-m_{\tilde l_k}^2)\right).
\ee
Moreover, for such small slepton decay widths one can also neglect
the interference between identical outgoing neutralinos, as we have 
checked numerically.

Integrating over the slepton momenta squared $p_3^2$ and $p_4^2$ and using the 
definitions
\begin{eqnarray}
 d\sigma_{ijkl} &=&(2\pi)^4\delta(p_1+p_2-p_3-p_4)
                   \frac{d^3p_3}{(2\pi)^32E_3}\frac{d^3p_4}{(2\pi)^32E_4}
                   \frac{M_{ij} M_{kl}^{*}}{2s}
\nonumber\\
dB_{ik} &=& (2\pi)^4\delta(p_3-p_5-p_7)
            \frac{d^3p_5}{(2\pi)^32E_5}\frac{d^3p_7}{(2\pi)^32E_7}
            \frac{C_{ik}}{2\overline{m\Gamma}_{ik}} M_i^- M_k^{-*}
\nonumber\\
dB_{jl} &=& (2\pi)^4\delta(p_4-p_6-p_8)
            \frac{d^3p_6}{(2\pi)^32E_6}\frac{d^3p_8}{(2\pi)^32E_8} 
            \frac{C_{jl}}{2\overline{m\Gamma}_{jl}} M_j^{\pm} M_l^{\pm*}
\end{eqnarray}
one can express the differential cross-sections in the form
\begin{equation}\label{shortform_dsigma}
d\sigma=\frac{1}{4} \sum_{ijkl} \sum_{p_3^2=m_{\tilde l_i}^2,m_{\tilde l_k}^2 
        \atop p_4^2=m_{\tilde l_j}^2,m_{\tilde l_l}^2}
        d\sigma_{ijkl}dB_{ik}dB_{jl}.
\end{equation}
As one expects on general grounds, for large mass differences, 
\(\Delta \tilde m^2_{ik} \gg \overline{m\Gamma}_{ik}\), 
the factors \(C_{ik}\) in (\ref{xsection}) approach \(\delta_{ik}\). 
Consequently, the coherent sum in
(\ref{shortform_dsigma}) reduces to an incoherent 
sum over 
products of production cross-sections times branching ratios:
\begin{equation}\label{sigma_decoherent}
  d\sigma=\sum_{ij} d\sigma(e^\pm e^- \to \tilde l_j^\pm \tilde l_i^-) 
dBr(\tilde l_i^- \to l_\alpha^- \tilde \chi^0_a) 
dBr(\tilde l_j^\pm \to l_\beta^\pm \tilde \chi^0_b).
\end{equation}

Previous studies often used only generic slepton masses and widths.
In the present analysis the slepton decay widths 
\(\Gamma_{\tilde l_i}\) are actually calculated for each of the 
mSUGRA benchmark scenarios, including the effects of LFV.
This is important for consistent phenomenological studies.

\section{Numerical results}

\subsection{SUSY and neutrino parameters}

Similarly as shown in \cite{Deppisch:2002vz} for the
radiative decays $l_\alpha \to l_\beta \gamma$, LFV in
$e^{\pm}e^-\rightarrow l_{\beta}^{\pm} l_{\alpha}^- 
\tilde{\chi}_b^0 \tilde{\chi}_a^0$ depends very sensitively 
on the SUSY scenario and on the neutrino masses and mixings.
The choice of models and the neutrino data 
are already described in some detail in the article mentioned above. 
Therefore, it suffices here to briefly specify the input parameters 
used in the following numerical analysis.
 
Again, we choose the mSUGRA benchmark scenarios proposed in 
\cite{Battaglia:2001zp} for linear collider studies.
In these set of  models the universal trilinear coupling 
parameter $A_0$ is assumed to vanish and, hence, LVF only 
occurs in the left-handed slepton sector as can be seen from
(\ref{eq:rnrges})-(\ref{eq:rnrges123}). 
For this reason, we focus on those scenarios which 
lead to sufficiently light sleptons, so that 
the production of slepton pairs containing a large
admixture of at least one left-handed   
$\tilde l_L$ component is possible at $\sqrt{s} = 500$~GeV.
The corresponding values of the mSUGRA parameters are 
listed in Tab.~\ref{mSUGRAscen}, together with some characteristic
predictions.  
 
\begin{table}[h!]
\begin{center}
\begin{tabular}{|c|c|c|c||c|c|c|}\hline
Scenario & $m_{1/2}$/GeV  & $m_{0}$/GeV & $\tan\beta$ & 
$m_{\tilde{l}_{6}}$/GeV &$\Gamma_{\tilde{l}_{6}}$/GeV & 
$m_{\tilde{\chi}_1^0}$/GeV  \\ \hline\hline
B & 250  & 100  & 10  & 208 & 0.32 & 98  \\ \hline
C & 400  & 90   & 10  & 292 & 0.22 & 164  \\ \hline
G & 375  & 120  & 20  & 292 & 0.41 & 154  \\ \hline
I & 350  & 180  & 35  & 313 & 1.03 & 143  \\ \hline
\end{tabular}
\end{center} 
\caption{\label{mSUGRAscen} Parameters of the relevant
mSUGRA benchmark scenarios (from \protect{\cite{Battaglia:2001zp}}). 
The sign of \(\mu\) is chosen to be positive and \(A_0\) is set to 
zero. Given are also the mass and total width of the heaviest
charged slepton and the mass of the lightest neutralino.}
\end{table}

Most likely, at the time when a linear collider will 
be in operation, more precise measurements of the
neutrino parameters will be available than today.
In order to simulate the expected improvement, 
we take the central values of the mass squared 
differences $\Delta m^2_{ij}=m_j^2-m_i^2$
and mixing angles $\theta_{ij}$ from a global fit to existing 
data \cite{Gonzalez-Garcia:2001sq} with errors that indicate 
the anticipated 90 \% C.L. intervals of running and proposed 
experiments as further explained in \cite{Deppisch:2002vz}: 
\begin{eqnarray}
&&\tan^2\theta_{23}=1.40^{+1.37}_{-0.66},~~
\tan^2\theta_{13}=0.005^{+0.001}_{-0.005},~~
\tan^2\theta_{12}=0.36^{+0.35}_{-0.16}, \label{nupar1}\\
&&\Delta m_{12}^2=3.30^{+0.3}_{-0.3}\cdot 10^{-5}\textrm{ eV}^2 ,~~
\Delta m_{23}^2=3.10^{+1.0}_{-1.0}\cdot 10^{-3}\textrm{ eV}^2.
\end{eqnarray}
The fit \cite{Gonzalez-Garcia:2001sq} is not the latest one, 
but still valid within the present uncertainties. It is taken 
for consistency with the results derived in \cite{Deppisch:2002vz} 
and used in the following.\footnote{In section 4.2 we will remark on
the slight changes in our results when using the central values of the 
neutrino parameters from one of the more recent fits \cite{Maltoni:2003da} 
instead of \cite{Gonzalez-Garcia:2001sq}.} 
The CP-violating Dirac phase $\delta$ is very 
difficult to measure and hence we allow it to vary in the full range
\begin{eqnarray}
&&\delta=0-2\pi.
\end{eqnarray}

Furthermore, we consider both hierarchical and quasi-degenerate 
spectra of light neutrinos. The case of neutrino masses that
are too light for an absolute mass measurement, is described by  
\begin{eqnarray}
&&m_{1}=0-0.03~{\rm eV},~~m_{2}= \sqrt{m_1^2 + \Delta
m^{2}_{12}},~~m_{3}=\sqrt{m_2^2 + \Delta m^{2}_{23}}
\label{hierspec}.
\end{eqnarray}
For $m_1 \ll m_2 \ll m_3$, the combination $Y_{\nu}^{\dagger}Y_{\nu}$, entering via 
(\ref{yukawprod}) the renormalization group correction (\ref{eq:rnrges})
to the slepton mass matrix, can be approximated by
\begin{equation}
\left(Y_{\nu}^{\dagger}Y_{\nu}\right)_{ij} \approx
\frac{M_{R}}{v^{2}\sin^{2}\beta}
\left(\sqrt{\Delta m^{2}_{12}}
U_{i2}U_{j2}^{*} + \sqrt{\Delta m_{23}^2}
U_{i3}U_{j3}^{*}\right). \label{llcorrectionhier}
\end{equation}

For the quasi-degenerate case we take  
\begin{eqnarray}
&&m_1=(0.3^{+0.11}_{-0.16})~\textrm{eV},
~~m_{2}\approx m_1+\frac{1}{2m_1}\Delta m^{2}_{12},~~m_{3}\approx
m_1+\frac{1}{2m_1}\Delta m^{2}_{23}.
\end{eqnarray}
Since in this case \(m_1 \gg \sqrt{\Delta m^{2}_{12}}\), \(\sqrt{\Delta m^{2}_{23}}\),
one can use the approximation
\begin{equation}
\left(Y_{\nu}^{\dagger}Y_{\nu}\right)_{ij} \approx
\frac{M_{R}}{v^{2}\sin^{2}\beta}
\left(m_{1}\delta_{ij}+\left(\frac{\Delta
m^{2}_{12}}{2 m_1}
U_{i2}U_{j2}^{*} + 
\frac{\Delta m^{2}_{23}}{2m_1}
U_{i3}U_{j3}^{*}\right)\right) \label{llcorrectiondeg}.
\end{equation}
The Majorana mass \(M_R\) is kept as a free 
parameter and is only constrained by the requirement that \(Y_\nu\)
should stay small enough for perturbation theory to hold.

\subsection{Predictions for \(e^+e^-\) and \(e^-e^-\) collisions}

Using the numerical input specified in the preceding subsection, 
we have calculated the cross-sections for
$e^+e^- \to \tilde{l}^+_j \tilde{l}_i^- \to \mu^+ e^-  
+2 \tilde{\chi}_{1}^{0}$ and $ \tau^+ \mu^-  +2 \tilde{\chi}_{1}^{0}$,
as well as the cross-section for $e^-e^- \to \tilde{l}_j^-\tilde{l}^-_i 
\to \mu^- e^- +2 \tilde{\chi}_{1}^{0}$.
In the SUSY scenarios under study R-parity is preserved,
and $\tilde{\chi}_{1}^{0}$ is the lightest SUSY particle and therefore stable.
Since $\tilde{\chi}_{1}^{0}$ is not detected, the observable final states 
consist of a pair of charged leptons with different flavor plus missing 
energy ($/\!\!\!\!E$). Thus one is dealing with a rather simple signal. 
Note that in the models defined in Tab.~\ref{mSUGRAscen} the heavier
neutralino states $\tilde{\chi}_{a}^{0}, \;a=2,3,4$ do not contribute 
to these signals: either they are too heavy and hence not or only 
very rarely produced in the slepton decays, or they are too light
in order to decay invisibly via 
$\tilde{\chi}_{a}^0 \to \tilde{\nu}\nu \to \tilde{\chi}_{1}^{0}\nu \nu$.
Other open channels such as $\tilde{\tau}\tau$ lead to more
complicated final states which are not considered here. 
As a rule of thumb, for an integrated luminosity of 1000~fb$^{-1}$, 
the goal at an $e^+e^-$ LC such as TESLA, 
the cross-section should be larger than $10^{-2}$ fb for a signal to become 
observable.

Fig.~\ref{fig:ep} shows the cross-sections 
for \(\mu^+ e^- + 2\tilde\chi^0_1\) and \(\tau^+\mu^- + 2\tilde\chi^0_1\) 
at \(\sqrt{s}=500\) GeV as a function of the 
right-handed Majorana mass scale \(M_R\) in 
scenario B which predicts the lightest sleptons. 
We have assumed a very light neutrino spectrum 
and have scattered the neutrino parameters in the error intervals
specified in (\ref{nupar1})-(\ref{hierspec}) according to Gaussian
distributions, except $m_1$ and $\delta$ 
for which we have taken flat distributions.
Each choice corresponds to one particular point in Fig.~\ref{fig:ep}.

A few comments concerning the main features of this plot are in order.
Similarly as the branching ratios \(Br(l_\alpha \to l_\beta \gamma)\) 
\cite{Deppisch:2002vz}, 
the cross-sections exhibit the typical proportionality
to \((Y_\nu^\dagger Y_\nu)^2 \propto M_R^2\). 
This behavior indicates that the signal 
cross-sections are dominated by Feynman graphs involving the 
lepton-flavor violating vertices in first order. The latter in turn implies 
that $e^+ e^- \to \tau^+ \mu^- + 2 \tilde{\chi}_{1}^{0}$
is dominantly a $s$-channel process and thus suppressed relative to 
$e^+ e^- \to \mu^+ e^- +2 \tilde{\chi}_{1}^{0}$
where the $s$- and $t$-channel contribute. 

Furthermore, in contrast to the branching ratios of the radiative decays,  
the cross-sections saturate for large \(M_R\). This saturation 
is seen most clearly in the \(\mu^+ e^-\) channel.
It can be understood by realizing that for large \(M_R\) the mass 
differences of the sleptons with a dominant left-handed component
become comparable to the corresponding slepton widths.
As already mentioned, in this case
the cross-sections can be approximated by the incoherent 
sum (\ref{sigma_decoherent}). Focussing on the $e\mu$ channel, the 
dominant contribution is determined by the \(\tilde{e}_L\)-\(\tilde{\mu}_L\) 
mixing angle or, more precisely, by 
\begin{equation}
\tan2\tilde{\theta}_{e\mu} \approx \frac{2 (m^2_{\tilde{l}_{L}})_{12}}
{(m^2_{\tilde{l}_{L}})_{11}-(m^2_{\tilde{l}_{L}})_{22}}\approx 
\frac{2(Y_{\nu}^\dagger Y_{\nu})_{12}}
{(Y_{\nu}^\dagger Y_{\nu})_{11}-(Y_{\nu}^\dagger Y_{\nu})_{22}},
\label{decomix}
\end{equation}
where the second equality follows from
(\ref{slepcorr_1}),
(\ref{left_handed_SSB}), and (\ref{eq:rnrges}) 
if the lepton masses are neglected. 
Using (\ref{llcorrectionhier}) one then sees that
$\tan2\tilde{\theta}_{e\mu}$ and thus also the cross-sections
are independent of \(M_R\). 
In the $\tau^+ \mu^-$ channel the saturation is less pronounced 
due to effects of the heavier $\tau$ mass which dominate the
denominator in the expression analogous to (\ref{decomix}),
except for very large values of \(M_R\). 
The finite $\tau$ mass is also  responsible 
for a suppression of all channels with a $\tau$ lepton 
in the final state relative to the $\mu e$ channel.

Finally, the impact of the neutrino uncertainties is weaker
in the \(\tau^+ \mu^-\) than in the \(\mu^+ e^-\) channel. 
This is not surprising, since the product of neutrino Yukawa couplings
$\left(Y_{\nu}^{\dagger}Y_{\nu}\right)_{23}$ relevant for 
\(\tau^+ \mu^-\) is mainly given by a single term depending on 
the large angle \(\theta_{23}\) and mass squared difference
\(\Delta m_{23}^2\), whereas $\left(Y_{\nu}^{\dagger}Y_{\nu}\right)_{12}$ 
relevant for \(\mu^+ e^-\) 
depends on two terms involving the small quantities \(\theta_{13}\)
and \(\Delta m_{12}^2\), respectively, as can be seen in
(\ref{llcorrectionhier}) and (\ref{llcorrectiondeg}) for $i \neq j$. 
Given the uncertainties in \(\theta_{13}\),
the two terms can become similar in magnitude and cancel
depending on the unknown phase $\delta$.  
The same holds for the $\tau e$ channel. Since this channel is unfavorable
because of the $\tau$ mass suppression pointed out above and the
comparably large MSSM background, we will concentrate on the  
\(\mu e\) and \(\tau \mu\) final states. 

The corresponding cross-sections for the higher cms 
energy $\sqrt{s}=800$~GeV are plotted in Fig.~\ref{fig:ep800}.
Here, the scenario I being unaccessible at 500 GeV is included. 
This figure illustrates the range of cross-sections which can be expected 
in the benchmark models specified in Tab. \ref{mSUGRAscen}. 
Together with Fig.~\ref{fig:ep} it also indicates the energy-dependence 
of the signals.

If the central values of the neutrino parameters from one of the
latest fits \cite{Maltoni:2003da} are used instead of (39) and (40), the most likely 
predictions for \(\sigma(e^+e^- \to \mu^+e^- + 2\tilde\chi^0_1)\) rise by 
about 30\%  as compared to the results shown on Fig. 3 and 4, while
the spread of the predictions shrinks slightly.
Furthermore, the cross sections for \(e^+e^- \to \tau^+\mu^- + 2\tilde\chi^0_1\)
decrease by about 20\% with the spread remaining unchanged.

We now turn to a particularly interesting subject, namely the 
correlations between the cross-sections 
$\sigma(e^\pm e^- \to \tilde{l}_j^{\pm}\tilde{l}^-_i
\to l_\beta^{\pm}l^-_\alpha\tilde{\chi}^0_b\tilde{\chi}^0_a)$ 
and the branching ratios $Br(l_\alpha \to l_\beta \gamma)$.
This correlation is illustrated in Fig.~\ref{fig:emu_lowhigh}
for $e^+e^- \to \mu^+e^- + 2\tilde\chi^0_1$ at $\sqrt{s}=500$~GeV
and $\mu \to e \gamma$. For \(M_R\lsim 10^{13}\)~GeV, 
corresponding roughly to \(Br(\mu\to e \gamma)\lsim 10^{-13}\)
\cite{Deppisch:2002vz}, this correlation is so accurate that the  
neutrino uncertainties drop out almost completely. 
Only for larger values of $M_R$ the correlation gets lost 
because of the saturation of the cross-sections explained 
above which sets in at different values of $M_R$ depending 
on the precise values of the neutrino parameters. 
Fig.~\ref{fig:mutau_lowhigh} shows the analogous relation between 
\(\sigma(e^+e^- \to \tau^+ \mu^- + 2 \tilde\chi_1^0)\) and  
\(Br(\tau\to \mu \gamma)\) which is even stronger.
Note that beyond the upper ends of the scatter plots in
Fig.~\ref{fig:mutau_lowhigh} the neutrino Yukawa couplings would 
become too large for perturbation theory to hold.

These correlations can be used to estimate the LC
cross-sections that are allowed 
by bounds on or measurements of the radiative decays.
Taking scenario B as an example, one can read off from 
Fig.~\ref{fig:emu_lowhigh} that the present bound 
\(Br(\mu\to e \gamma) <1.2\cdot 10^{-11}\) \cite{Hagiwara:pw} 
implies \(\sigma(e^+e^- \to \mu^+ e^- + 2 \tilde\chi_1^0)< 0.3~{\rm fb}\), 
while a measurement of \(Br(\mu\to e \gamma) \approx 10^{-14}\)
by the new experiment at PSI \cite{barkov}
would predict \(\sigma(e^+e^- \to \mu^+ e^- + 2 \tilde\chi_1^0) \approx 
3\cdot 10^{-4}~{\rm fb}\). In other words, 
if $\mu \to e \gamma$ will not be detected at PSI, one does not expect an
observable $\mu e$ signal at a 500 GeV LC either. In model C,
the above branching ratios imply 
\(\sigma(e^+e^- \to \mu^+ e^- + 2 \tilde\chi_1^0)< 1~{\rm fb}\) and
$\approx 6\cdot 10^{-3}~{\rm fb}$, respectively.

Analogously, from Fig.~\ref{fig:mutau_lowhigh} one concludes 
that the present bound 
\(Br(\tau\to \mu \gamma)<6 \cdot 10^{-7}\) \cite{Inami:2002us}
provides the weak constraints
$\sigma(e^+e^- \to \tau^+ \mu^- + 2 \tilde\chi_1^0) < 10$~fb 
at \(\sqrt{s}=800\)~GeV for model B, C, and G, and $<2$~fb for model I.
More importantly, also the sensitivity goal 
\(Br(\tau\to \mu \gamma) \approx 10^{-9}\)  
of future searches \cite{superkekb} will not rule out sizable 
LC cross-sections, namely 
\(\sigma(e^+e^- \to \tau^+ \mu^- + 2 \tilde\chi_1^0) \approx 6\;(0.2)\)~fb
in scenario C (B and G).
In fact, in the $\tau \mu$ channel LC experiments may reach farther than 
the $\tau \to \mu \gamma$ experiments planned in the future. 
At any rate, here we have
a nice example for the complementarity of low and high-energy searches.

In principle, one has such correlations also in channels which differ 
in flavor. This is exemplified in Fig.~\ref{fig:mutau_emu_lowhigh}
for \(\sigma(e^+e^- \to \tau^+ \mu^- + 2 \tilde\chi_1^0)\) 
and \(Br(\mu\to e \gamma)\). However, because of the
different flavor-violating couplings involved this correlation
suffers considerably from uncertainties in the neutrino sector.
Nevertheless, the experimental bound 
\(Br(\mu\to e \gamma)<1.2 \cdot 10^{-11}\)
yields a stronger constraint on model I than the one obtained
from Fig.~\ref{fig:mutau_lowhigh}, excluding 
$\sigma(\tau^+ \mu^- +2 \tilde\chi_1^0) > {\rm few}\cdot 10^{-2}$~fb  
at \(\sqrt{s}=800\)~GeV already today.

Finally, it is interesting to look at 
the relation between the two cross-sections
\(\sigma(e^+e^- \to \tau^+ \mu^- + 2 \tilde\chi_1^0)\)
and \(\sigma(e^+e^- \to \mu^+ e^- + 2 \tilde\chi_1^0)\)
themselves. This relation is displayed in  
Fig.~\ref{fig:highhigh800} for 800 GeV and scenario C.
Unfortunately, it is washed out by the uncertainties in the neutrino 
parameters. Nevertheless, a measurement in the $\mu e $ channel 
would give a useful lower bound in the $\tau \mu$ channel.

The use of the latest central values of the neutrino parameters from \cite{Maltoni:2003da}
instead of (39) and (40) does not change the correlations shown in 
Fig. 5 and 6 significantly. On the other hand, the lower boundaries of 
the correlations displayed in Fig. 7 and 8 are slightly shifted to the 
right.

For completeness, we have also studied the prospects  
for \(e^-e^-\) collisions. Fig.~\ref{fig:ee800} shows
\(\sigma(e^-e^- \to \mu^- e^- + 2 \tilde\chi_1^0)\)
and \(\sigma(e^-e^- \to \tau^- \mu^- + 2 \tilde\chi_1^0)\)
for \(\sqrt{s}=800\)~GeV and model C. 
The strong suppression of the \(\tau^-\mu^-\) final state
results from the fact that at least two flavor-violating couplings 
are required because of the absence of the $s$-channel 
(see Fig. \ref{e-e-_diags}).
The \(\tau^- e^-\) final state is suppressed for similar reasons
as the  \(\tau^+ e^-\) channel in $e^+e^-$ collisions.
Thus, \(\mu^-e^-\) remains as the only promising channel. 
The correlation between 
\(\sigma(e^-e^- \to \mu^- e^- + 2 \tilde\chi_1^0)\) 
and \(Br(\mu\to e \gamma)\) is plotted in
Fig.~\ref{fig:emu_lowhigh_ee} for the same scenario C. 
We see that the cross-section permitted 
by the present bound on $\mu\to e \gamma$
lies in the range of 1-10~fb which would make the detection
rather easy.
In this figure, we also show the prediction for a degenerate neutrino spectrum.
Generally, both the LC cross-sections  
as well as the radiative branching ratios tend to be smaller
for degenerate than for hierarchical neutrinos. 
This reduction comes from the 
suppression of the lepton-flavor violating terms in (\ref{llcorrectiondeg})
by the neutrino mass $m_1$ which sets the absolute mass scale. 
However, the correlation between cross-sections and branching ratios 
is found to be more or less independent of the absolute neutrino mass scale.
The predictions shown in Fig. 9 would change only slightly, 
if the central values of the neutrino parameters from \cite{Maltoni:2003da}
are used instead of (39) and (40), similarly as described 
above for \(e^+e^-\) scattering. Again, the correlation depicted 
in Fig. 10 remains unaffected.

\section{Background processes}

After having discussed the lepton-flavor violating 
signals leading to the channels 
$l_{\beta}l_{\alpha}+/\!\!\!\!E$
we turn now to the lepton-flavor conserving
standard model (SM) and SUSY background. 
For the dominant SM reactions we also illustrate the efficiency 
of angular and energy cuts in reducing this background. 
The SUSY background processes are 
calculated in the respective mSUGRA benchmark scenarios
without cuts.
We are going to assume that at the time these searches 
for LFV will be performed, SUSY has already been discovered
and the lighter sparticle masses and main decay channels are known.
In this case, one will be able to design very specific cuts optimized
to each channel of interest.  
Under this assumption our estimates indicate that it should be possible 
to reduce the SM and SUSY background sufficiently
and detect some of the signals considered, provided the
Majorana scale $M_R$ is not much smaller than $10^{13}$~GeV. However,  
more conclusive statements on the feasibility to discriminate signal 
from background require detailed simulations  
which are beyond the scope of this paper.

\subsection{Background in \(e^+e^-\)}

In the SM, the dominant background in the channels
$e^+ e^- \to \tilde{l}_j^+\tilde{l}^-_i\to 
l_\beta^+l^-_\alpha + 2\tilde{\chi}^0_1$  
is produced by the following 
lepton-flavor conserving processes:
\begin{enumerate}[B1)]
\item $e^+e^-\rightarrow W^+W^-\rightarrow  
l_\beta^+\nu_\beta l_\alpha^-\bar{\nu}_\alpha$ \qquad \((\alpha\neq \beta)\),
\item $e^+e^-\rightarrow W^+ e^-\bar{\nu}_e\rightarrow  
l_\beta^+\nu_\beta e^-\bar{\nu}_e$ \qquad\( (\beta\neq 1)\), 
\item $e^+e^-\rightarrow \tau^+\tau^-\rightarrow 
\tau^+\nu_{\tau}l^-_\alpha\bar{\nu}_\alpha$ \qquad\( (\alpha=1,2)\).
\end{enumerate}
These have been estimated with the help of the COMPHEP program package
\cite{Pukhov:1999gg}, including the following kinematical cuts: 
\begin{enumerate}[i)]
\item beam-pipe cut: 
$|\cos(e^\pm,l^\pm_{\beta,\alpha})|<0.985$ ($10^{\circ}$),
\item lepton-energy cut: 
\(E_{min} \leq E_l \leq E_{max}\),  
\item missing-energy cut: 
\(2m_{\tilde{\chi}_1^0}\leq /\!\!\!\!E\leq\sqrt{s}-2E_{min}\).
\end{enumerate}
The angular cut i) partially eliminates the large contribution from $t$-channel 
photon exchange (B2), and also small-angle $W$ (B1) and $\tau$ (B3) production. 
Since the angular distributions of the decay leptons 
from the heavy sleptons are relatively flat, the signal rates are thereby reduced 
by less than 10\%. 
Then, knowing the slepton spectrum and the LSP mass, one can impose the
cuts (ii) and (iii) requiring the lepton energies and the total missing energy 
to lie in the intervals corresponding to the decay 
$\tilde{l}\rightarrow l \tilde{\chi}^0_1$. Since one is dealing with 
a coherent process, the intermediate slepton flavor is not known. Therefore,
$E_{min}$ ($E_{max}$) is defined as the minimum (maximum) 
of the decay energies kinematically allowed for any slepton flavor:
\beq 
E_{min} = \min\left(\frac{m_{\tilde{l}}^2-m^2_{\tilde{\chi}_1^0}}
{2(E_{\tilde{l}} + p_{\tilde{l}})}\right), \; 
E_{max} = \max\left(\frac{m_{\tilde{l}}^2-m^2_{\tilde{\chi}_1^0}}
{2(E_{\tilde{l}} - p_{\tilde{l}})}\right).
\eeq

\begin{table}[h!]
\begin{center}
\begin{tabular}{|c|c|c|c||c|c|c|} \hline
Final state &$\sqrt{s}$/GeV & Cut & $\sigma/fb$ & $\sqrt{s}$/GeV & Cut & $\sigma/fb$   \\ \hline
$\mu^+e^-\bar{\nu}_e\nu_{\mu}$& 500 & i            & 105.5 & 800 & i              & 85.1    \\ \hline
                              &    & i+ii+iii, B     & 29.6  &     & i+ii+iii, B  & 27.5    \\ \hline
                              &    & i+ii+iii, C     & 9.2   &     & i+ii+iii, C  & 19.0    \\ \hline
                              &    & i+ii+iii, G     & 11.6  &     & i+ii+iii, G  & 20.8    \\ \hline
                              &    & i+ii+iii, I     & 10.8  &     & i+ii+iii, I  & 24.2    \\ \hline
\hline
Final state &$\sqrt{s}$/GeV & Cut & $\sigma/fb$ & $\sqrt{s}$/GeV & Cut & $\sigma/fb$ \\ \hline
$\tau^+\mu^-\bar{\nu}_{\mu}\nu_{\tau}$ & 500 &  i        & 58.8  & 800  & i             & 22.5  \\ \hline    
                                       &    & i+ii+iii, B  & 11.1  &      &i+ii+iii, B   & 6.0   \\ \hline
                                       &    & i+ii+iii, C  & 3.1   &      &i+ii+iii, C   & 3.5   \\ \hline 
                                       &    & i+ii+iii, G  & 3.5   &      &i+ii+iii, G   & 3.9   \\ \hline
                                       &    & i+ii+iii, I  & 3.6   &      &i+ii+iii, I   & 4.8   \\ \hline
\end{tabular}
\end{center}
\caption{Cross-sections for the 
SM background in $e^+e^-$ collisions including cuts as explained in the text.}
\label{SM_back_cross}
\end{table}

The SM background cross-sections that remain after these cuts 
are summarized in Tab.~\ref{SM_back_cross}. If one requires a signal
to background ratio $S/\sqrt B = 2$ and assumes a typical signal cross-section
of 0.1~fb, one can afford a background of about 2~fb. 
The additional suppression may be achieved by applying the standard selectron 
selection cuts \cite{Becker93} on the acoplanarity, 
lepton polar angle and missing transverse momentum. 
It has been shown that in this way the SM background 
to slepton-pair production can be reduced to about 2-3~fb
at $\sqrt{s}=500$~GeV, 
while the signal cross-section shrinks only by a factor 3.
It should be noted that suppression of the $W$-pair background by right-handed electron 
polarization is not an option here, since in the mSUGRA scenarios considered  
LFV occurs only in the left-handed slepton sector, and thus right-handed polarization 
would also kill the signal.

In the MSSM, the main background processes are

\begin{itemize}
\item [B4)] $e^+e^-\rightarrow \tilde{\nu}_\alpha\bar{\tilde{\nu}}_{\alpha}$, 
\item [B5)] $e^+e^-\rightarrow \tilde{l}_\alpha^+\tilde{l}_\alpha^-$, 
\item [B6)] $e^+e^-\rightarrow \tilde{\chi}_b^+\tilde{\chi}_a^-$,
\item [B7)] $e^+e^-\rightarrow \tilde{\chi}_{b}^+ e^-\bar{\tilde{\nu}}_{e}$,
\end{itemize}

with the sparticles decaying like
$\tilde{\nu}_\alpha \to \nu_\alpha \tilde{\chi}_{1}^{0}$ and 
$l^-_\alpha \tilde{\chi}_{1}^{+}$,
$\tilde{\chi}_{1}^{+} \to l_{\alpha}^+ \tilde{\nu}_{\alpha}$ and 
$\tilde{l}_{\alpha}^+\nu_{\alpha}$,
$\tilde{l}_{\alpha}^+ \to l_{\alpha}^+ \tilde{\chi}_1^0$ and 
$\bar{\nu}_{\alpha}\tilde{\chi}^+_1$,
and many other channels, depending on the given sparticle spectrum.

The numerical estimates for the MSSM background listed
in Tab.~\ref{MSSM_back_cross} are again obtained 
with the help of COMPHEP. Here, no cuts are included and the  
cross-sections for the individual production channels are added
incoherently.
We find that the MSSM background to  
$\mu^+e^-+\;/\!\!\!\!E$ is very small, below 0.2~fb in all scenarios 
of Tab.~\ref{mSUGRAscen} except for model I and C, where it amounts 
to 0.4 and 5~fb, respectively, at $\sqrt{s}=800$~GeV.
With 2-7~fb the \(\tau^+\mu^-\) background is considerably bigger.
Also shown in Tab.~\ref{MSSM_back_cross} is the MSSM
background to $\tau^+ e^-+\;/\!\!\!\!E$, which can contribute to the \(\mu^+e^-\)
channel via the decay $\tau^+ \rightarrow \mu^+\nu_{\mu}\bar{\nu}_{\tau}$.
If $\tilde{\tau}_1$ and $\tilde{\chi}^+_1$ are very light, like in scenarios B 
and I, this background can be as large as 100~fb.
The charginos originate mainly from selectron or 
$\tilde{\nu}_e$ pair production and decay almost exclusively into staus, 
$\tilde{\chi}^-_{1}\rightarrow \tilde{\tau}_1^-\bar{\nu}_{\tau}$.
However, such events typically contain two neutrinos in addition to the
two LSPs which are also present in the signal events. Thus, after $\tau$ decay 
one has altogether
six invisible particles instead of two, which should allow to discriminate the signal 
in $\mu^+e^-+\;/\!\!\!\!E$ also from this potentially 
dangerous MSSM background 
by cutting on various distributions.

\begin{table}
\begin{center}
\begin{tabular}{|c|c|c|c|c|c|c|c|}\hline
Final state & Production  & \multicolumn{2}{|c|}{B} &  C & \multicolumn{2}{|c|}{G} & I  \\ \hline 
$\mu^+e^-+/\!\!\!\!E$ &                                        & 500     & 800        & 800   
& 500     & 800   & 800 \\\hline
                    & $\tilde{l}_i^+\tilde{l}_i^-$             & 0       &  0         & 0     
&0.02     & 0.06  &  0.1  \\\hline
                    & $\bar{\tilde{\nu}}_i\tilde{\nu}_i$       & 0.01    &  0.01      & 0     
&-        & 0     & 0.3   \\\hline
                    & $\tilde{\chi}_b^+\tilde{\chi}_a^-$       & 0       &  0.07      & 2.5  
&-        & 0.05  &0.01    \\\hline
                    & $\tilde{\chi}^+_1e^-\bar{\tilde{\nu}}_e$ & 0       &  0.04      & 2.3   
&-        & 0.07  & 0   \\\hline
                    &   Sum                                    & 0.01    &  0.12      & 4.8   
&0.02     & 0.18  & 0.41   \\\hline\hline
$\tau^+\mu^-+/\!\!\!\!E$ &                                     & 500     & 800        &  800  
& 500     &800    & 800 \\\hline
                     & $\tilde{l}_i^+\tilde{l}_i^-$            &   3.5   & 4.4        &  0    
& 0       &  0.38 & 2.0   \\\hline
                     & $\bar{\tilde{\nu}}_i\tilde{\nu}_i$      &   1.8   & 1.6        &  0.09 
& -       &  0.19 & 1.2   \\\hline
                     & $\tilde{\chi}_b^+\tilde{\chi}_a^-$      &  0.02   &  1.3       &  5.9  
& -       &  1.7  & 0.8   \\\hline
                     &  Sum                                    &  5.3    & 7.3        &  6.0  
& 0       & 2.3   & 4.0   \\\hline \hline
$\tau^+e^-+/\!\!\!\!E$ &                                       & 500     & 800        & 800  
 & 500     & 800   & 800 \\\hline
                    & $\tilde{l}_i^+\tilde{l}_i^-$             & 38.2    & 37.5       & 0    
 & 0.8     & 2.0   &16.2    \\\hline
                    & $\bar{\tilde{\nu}}_i\tilde{\nu}_i$       & 60.0    & 81.5       &  0.09 
& -       & 0.2   & 38.7   \\\hline
                    & $\tilde{\chi}_b^+\tilde{\chi}_a^-$       & 0       &  0.8       &  5.9  
& -       & 1.7   & 0.5   \\\hline
                    & $\tilde{\chi}^+_1e^-\bar{\tilde{\nu}}_e$ & 0       & 0.04       &  1.7  
& -        & 0.05  & 0   \\\hline
                    & Sum                                      & 98.2    & 119.8      & 7.7   
&0.8      & 4.0   & 55.4   \\\hline
\end{tabular}
\end{center}
\caption{Cross-sections in fb for the MSSM background 
in $e^+e^-$ collisions at \(\sqrt{s}=500\)~GeV and 800~GeV. 
The second column specifies the important production channels. 
In scenarios C and I the cross-sections are below 0.01~fb at $\sqrt{s}=500$~GeV
and are therefore omitted. Otherwise, cross-sections below 0.01~fb are denoted by 0.  
Kinematically forbidden channels are marked by a hyphen.}
\label{MSSM_back_cross}
\end{table}

\subsection{Background in \(e^-e^-\)}
In seesaw models the process 
$e^-e^- \to W^-W^- \to l^-_\beta l^-_\alpha \bar{\nu}_\beta \bar{\nu}_{\alpha}$ 
(\(\alpha,\beta \neq 1\)) via neutrino exchange
is highly suppressed due to the very small admixture of the
heavy Majorana neutrinos in the light neutrino eigenstates \cite{Gluza:1995ix}.
The analogous argument holds for the SUSY version of this process, i.e.
chargino-pair production via $t$-channel sneutrino exchange.
The main background processes leading to the final state 
$e^-e^- \to \mu^-e^-+\;/\!\!\!\!E$ are 
single $W^-$ production (cf. B2), 
charged slepton-pair production (cf. B5), and
single chargino production (cf. B7).

The SM background $e^-\mu^-\nu_e\bar{\nu}_{\mu}$ 
originating mainly from single $W^-$ production 
is estimated in  Tab.~\ref{eeSM_back_cross}
including the cuts (i)-(iii) explained in the preceding subsection. 
As can be seen, the cross-sections
are bigger and the cuts are less efficient than in the 
$e^+e^-$ case. Moreover, contrary to what we have found for $e^+e^-$ 
collisions,
the situation at \(\sqrt{s}=800\)~GeV is somewhat worse than at 500~GeV
due to the increase of single \(W^-\) production with energy. 

\begin{table}
\begin{center}
\begin{tabular}{|c|c|c||c|c|c|c|} \hline
Final state &$\sqrt{s}$/GeV & Cut & $\sigma/fb$ & $\sqrt{s}$/GeV & Cut & $\sigma/fb$ \\ \hline
 $e^-\mu^-\nu_e\bar{\nu}_\mu$    & 500 & i  & 122.8 & 800                & i                 & 175.7   \\ \hline
                                 &     & i+ii+iii, B          & 55.5  &  & i+ii+iii, B       & 79.5    \\ \hline
                                 &     & i+ii+iii, C          & 13.9  &  & i+ii+iii, C       & 55.2    \\ \hline
                                 &     & i+ii+iii, G          & 17.8  &  & i+ii+iii, G       & 61.1    \\ \hline
                                 &     & i+ii+iii, I          & 19.4  &  & i+ii+iii, I       & 70.8    \\ \hline
\end{tabular}
\end{center}
\caption{Cross-sections for SM background processes in $e^-e^-$ collisions including cuts as explained in the text.}
\label{eeSM_back_cross}
\end{table}

The total cross-sections for the MSSM background processes 
are summarized in Tab.~\ref{eeMSSM_back_cross}. 
In all four scenarios the MSSM background to $\mu^- e^- +/\!\!\!\!E$ 
is below 1~fb, while the background to $\tau^- e^- +/\!\!\!\!E$ in 
scenarios B and I is of the order of 100~fb. This is similar to our 
findings for \(e^+e^-\) collisions. 
Despite the larger background, the signal to background ratio tends 
to be slightly more favorable
in $e^-e^-$ collisions than in $e^+e^-$. For example, for scenario C and
$\sqrt{s}=800$~GeV one finds roughly $S/\sqrt{B}(e^-e^-) \approx 2~
S/\sqrt{B}(e^+e^-)$, assuming an integrated luminosity of 250 and 1000
fb$^{-1}$, respectively. 

\begin{table}
\begin{center}
\begin{tabular}{|c|c|c|c|c|c|c|c|c|c|}\hline
Final state & Production  & \multicolumn{2}{|c|}{B}&  C &\multicolumn{2}{|c|}{G} & I  \\ \hline 
$e^-\mu^-+/\!\!\!\!E$  &               & 500   & 800   & 800  & 500  &800   &800    \\\hline
& $\tilde{e}_i^-\tilde{e}_i^-$         & 0.01  & 0.01  & 0    & 0.01 & 0.2  & 0.4   \\\hline      
& $e^-\tilde{\nu}_e\tilde{\chi}^-_1$   & 0     & 0.01  & 0.4  & -    & 0.1  &0      \\\hline
&  Sum                                 & 0.01  & 0.02  & 0.4  & 0.01 & 0.3  &0.4    \\\hline
\hline
$e^-\tau^-+/\!\!\!\!E$ &               & 500   & 800   & 800  & 500  & 800  &800 \\\hline
& $\tilde{e}_i^-\tilde{e}_i^-$         & 140.8 & 102.3 & 0    & 0.2  & 8.1  &  60.0   \\\hline
& $e^-\tilde{\nu}_e\tilde{\chi}^-_1$   & 0     & 0.01  & 0.4  & -    & 0.07 &0    \\\hline
&  Sum                                 & 140.8 & 102.3 & 0.4  & 0.2  & 8.2  &60.0    \\\hline
\end{tabular}
\end{center}
\caption{Cross-sections in fb for MSSM background processes in $e^-e^-$ collisions 
at \(\sqrt{s}=500\)~GeV and 800~GeV
as in Tab.~\ref{MSSM_back_cross}.}
\label{eeMSSM_back_cross}
\end{table}

\section{Conclusions}
We have studied the lepton flavor-violating processes $e^\pm e^- \to 
\sum_{i,j}\tilde{l}_j^{\pm}\tilde{l}^-_i\to 
l_\beta^{\pm}l^-_\alpha\tilde{\chi}^0_b\tilde{\chi}^0_a$  at a future
linear $e^+ e^-$ collider. 
As a theoretical framework we have chosen mSUGRA benchmark scenarios 
furnished with the seesaw mechanism assuming a Majorana mass scale 
in the range \(10^{11}~{\rm GeV}\le M_R \le 10^{15}\)~GeV. 
Our analysis shows that for center-of-mass energies of 500-800~GeV, the 
signal cross-sections may be as large as 1-10~fb. 
The detailed predictions depend strongly on the SUSY parameters 
and on the neutrino masses and mixing. 
Consequently, they are also strongly affected by uncertainties 
in the neutrino data.

On the other hand, the correlations between the high-energy 
cross-sections and the branching ratios for the corresponding rare radiative
decays are expected to be less influenced by 
experimental errors in neutrino parameters. 
We have therefore investigated these correlations very thoroughly 
and find that they are the stronger, the smaller the branching
ratios for the radiative decays, that is the lower $M_R$. 
At $M_R$ smaller than \(10^{13}\)~GeV the neutrino uncertainties 
play no role at all.  
The present bounds on $\mu \rightarrow e \gamma$ and 
$\tau \rightarrow \mu \gamma$ still allow sizable signals at a LC.
If $\mu \rightarrow e \gamma$
will not be observed at the new PSI experiment, 
the cross-section for $e^+e^- \to 
\sum_{i,j}\tilde{l}_j^+\tilde{l}^-_i\to 
\mu^+ e^- +2 \tilde{\chi}^0_1$ is predicted not to exceed 
0.02 fb at $\sqrt s = 500$~GeV. On the other hand, 
$Br(\tau \to \mu \gamma) < 10^{-9}$ alone would still be compatible
with a cross-section of order 1 fb for 
$e^+e^- \to \sum_{i,j}\tilde{l}_j^+\tilde{l}^-_i\to 
\tau^+ \mu^- +2\tilde{\chi}^0_1$, whereas $Br(\mu \to e \gamma)<10^{-14}$
would be in conflict with such a large cross-section.

Finally, we have estimated the SM and MSSM
background. 
With only a beam pipe cut of 10 degrees included, the
SM background at 500 to 800~GeV amounts to 80 to 100~fb 
in the $\mu^+ e^-$ channel and to 20 to 60~fb in the 
$\tau^+\mu^-$ channel. 
The corresponding MSSM background varies between 0.01 and 10~fb 
depending on the SUSY scenario, the final state and the beam energy. 
An exceptionally large MSSM background of 50 to 100~fb is found in 
scenarios B and I in the channel $\tau^+ e^-$ which also contributes to the 
$\mu^+ e^-$ 
channel through leptonic $\tau$ decay. 
Since one can assume that the relevant sparticle spectra are
known at the time such experiments can be done, it appears possible
to sufficiently suppress the MSSM background by carefully designed
cuts on decay energies, missing energy, angular distributions and other
quantities. However, conclusive feasibility studies require a detailed 
Monte Carlo simulation.

The overall discovery potential for LFV at a LC is slightly increased by
performing searches in $e^-e^-$ collisions in addition to $e^+e^-$ collisions.

\section*{Acknowledgements}   
We thank W. Porod for useful discussions.
This work was supported by the Bundesministerium f\"ur 
Bildung und Forschung (BMBF, Bonn, Germany) under the contract number 
05HT1WWA2 and (FD) by Spanish grant BFM2002-00345, by the European Commission
RTN grant HPRN-CT-2000-00148 and European Commission Research 
Training Site grant 
HPMT-CT-2000-00124.

\clearpage

\begin{center}
\begin{figure}[t!]
\( \sum_{i,j} \) 
\begin{picture}(235,50)(0,48)
                           \ArrowLine(50,50)(10,90)        \Text(12,100)[r]{$e^{+}(p_2)$}
                           \ArrowLine(10,10)(50,50)        \Text(12,0)[r]{$e^{-}(p_1)$}              \Vertex(50,50){2}
                           \Photon(50,50)(100,50){3}{5}           \Text(75,54)[b]{\(\gamma,Z\)} \Vertex(100,50){2}
                           \DashArrowLine(130,80)(100,50){5}  \Text(108,70)[b]{\(\tilde{l}^{+}_{j}(p_4)\)}              \Vertex(130,80){2}
                           \ArrowLine(160,100)(130,80)        \Text(162,100)[l]{\(l_{\beta}^{+}(p_6)\)}
                           \ArrowLine(130,80)(160,60)        \Text(162,60)[l]{\(\tilde{\chi}^{0}_{b}(p_8)\)}
                           \DashArrowLine(100,50)(130,20){5}  \Text(108,30)[t]{\(\tilde{l}_{i}^{-}(p_3)\)}          \Vertex(130,20){2} 
                           \ArrowLine(130,20)(160,40)        \Text(162,40)[l]{\(l^{-}_{\alpha}(p_5)\)}
                           \ArrowLine(160,0)(130,20)        \Text(162,0)[l]{\(\tilde{\chi}_{a}^{0}(p_7)\)}
                         \end{picture}\(+\sum_{c,i,j}\)
                         \begin{picture}(100,50)(5,48)
                           \ArrowLine(20,80)(0,100)   \Text(0,105)[r]{$e^{+}$}
                           \ArrowLine(0,0)(20,20)     \Text(0,-5)[r]{$e^{-}$} 
                           \ArrowLine(20,20)(20,80)   \Text(22,50)[l]{\(\tilde\chi_c^{0}\)} \Vertex(20,20){2} \Vertex(20,80){2}
                           \DashArrowLine(40,100)(20,80){5} \Text(28,93.5)[b]{\(\tilde{l}_{j}^{+}\)} \Vertex(40,100){2}
                           \ArrowLine(70,100)(40,100)  \Text(72,100)[l]{\(l_{\beta}^{+}\)}
                           \ArrowLine(40,100)(70,70) \Text(70,70)[l]{\(\tilde{\chi}_{b}^{0}\)}
                           \DashArrowLine(20,20)(40,0){5} \Text(28,5)[t]{\(\tilde{l}^{-}_{i}\)}    \Vertex(40,0){2}
                           \ArrowLine(40,0)(70,30)  \Text(70,0)[l]{\(\tilde{\chi}_{a}^{0}\)}
                           \ArrowLine(70,0)(40,0) \Text(70,30)[l]{\(l^{-}_{\alpha}\)}
\end{picture}
\vspace{1.9cm}
\\
\noindent
\caption{Feynman diagrams for $e^+e^-\to \tilde{l}_j^+\tilde{l}^-_i 
\to l_\beta^+
l^-_\alpha\tilde{\chi}^0_b\tilde{\chi}^0_a$.
The arrows on scalar lines refer to the lepton number flow.
}\label{e+e-_diags}
\vspace*{1cm}
\end{figure}
\end{center}


\begin{center}
\begin{figure}[t!]
\begin{picture}(400,150)(-200,-75)
\Text(12,0)[]{$+\sum_{c,i,j}$}
\Vertex(80,30){2}
\Vertex(80,-30){2}
\ArrowLine(50,60)(80,30)
\Text(47,65)[]{$e^{-}$}
\ArrowLine(50,-60)(80,-30)
\Text(44,-65)[]{$e^{-}$}
\ArrowLine(80,30)(80,-30)
\Text(68,0)[]{$\tilde{\chi}_{c}^{0}$}
\DashArrowLine(80,30)(110,-60){5}
\Text(115,32.5)[]{$\tilde{l}_{j}^{-} $}
\DashArrowLine(80,-30)(110,60){5}
\Text(115,-32.5)[]{$\tilde{l}_{i}^{-}$}
\ArrowLine(110,60)(140,80)
\Text(148,83)[]{$l^{-}_{\beta}$}
\Vertex(110,60){2}
\ArrowLine(140,-80)(110,-60)
\Text(148,-83)[]{$\tilde{\chi}_{a}^{0}$}
\Vertex(110,-60){2}
\Text(-190,0)[]{$\sum_{c,i,j}$}
\Vertex(-120,30){2}
\ArrowLine(110,-60)(140,-40)
\Text(148,-43)[]{$l^{-}_{\alpha}$}
\Text(148,43)[]{$\tilde{\chi}^{0}_{b}$}
\ArrowLine(140,40)(110,60)
\Vertex(-120,-30){2}
\ArrowLine(-150,60)(-120,30)
\Text(-166,68)[]{$e^{-}(p_2)$}
\ArrowLine(-150,-60)(-120,-30)
\Text(-166,-68)[]{$e^{-}(p_1)$}
\ArrowLine(-120,30)(-120,-30)
\Text(-133,0)[]{$\tilde{\chi}_{c}^{0}$}
\DashArrowLine(-120,30)(-90,60){5}
\Text(-95,31)[]{$\tilde{l}_{j}^{-}(p_4)$}
\DashArrowLine(-120,-30)(-90,-60){5}
\Text(-95,-31)[]{$ \tilde{l}_{i}^{-}(p_3)$}
\Vertex(-90,60){2}
\Vertex(-90,-60){2}
\ArrowLine(-90,60)(-60,80)
\Text(-41,83)[]{$l_{\beta}^{-}(p_6)$}
\ArrowLine(-60,-80)(-90,-60)
\Text(-41,-83)[]{$\tilde{\chi}_{a}^{0}(p_7)$}
\ArrowLine(-60,40)(-90,60)
\Text(-41,37)[]{$\tilde{\chi}^{0}_{b}(p_8)$}
\ArrowLine(-90,-60)(-60,-40)
\Text(-41,-37)[]{$l^{-}_{\alpha}(p_5)$}
\end{picture}
\vspace{0.5cm}
\\
\noindent
\caption{Feynman diagrams for $e^-e^- \to \tilde{l}_j^-\tilde{l}^-_i \to 
         l_\beta^-l^-_\alpha\tilde{\chi}^0_b\tilde{\chi}^0_a$.
The arrows on scalar lines refer to the lepton number flow.}
\label{e-e-_diags}
\vspace*{1cm}
\end{figure}
\end{center}

\begin{figure}[t]
\centering
\includegraphics[clip]{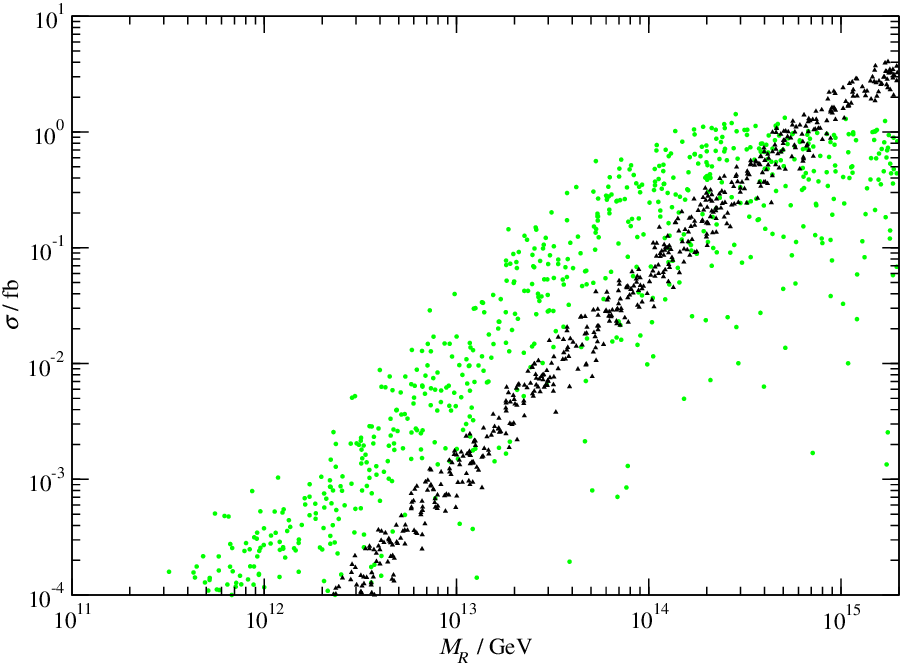}
     \caption{Cross-sections at \(\sqrt{s}=500\) GeV
              for \(e^+e^- \to \mu^+e^- +2\tilde\chi_1^0\) (circles)
              and  \(e^+e^- \to \tau^+\mu^- +  2\tilde\chi_1^0\) (triangles) 
              in scenario B  for the case of very light neutrinos.}
     \label{fig:ep}
\end{figure}

\begin{figure}[t]
\centering
\includegraphics[clip]{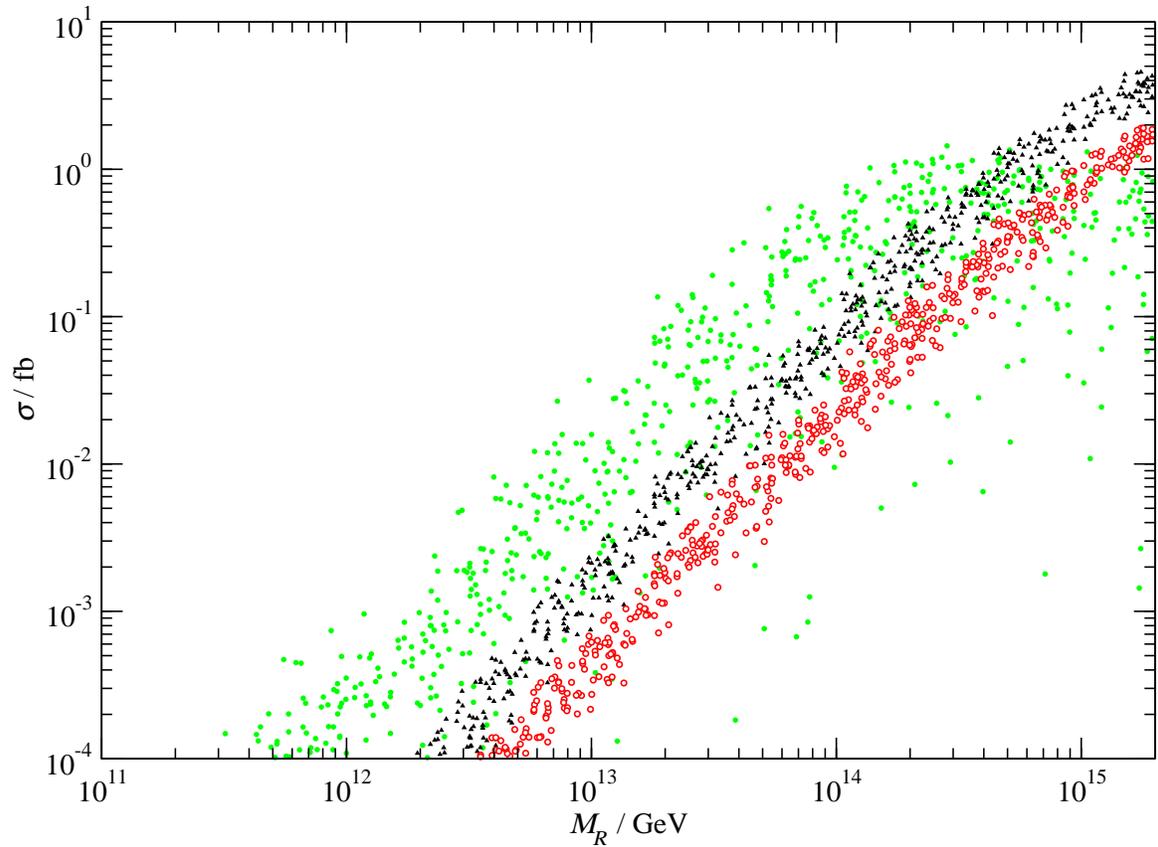}
     \caption{Cross-sections  at \(\sqrt{s}=800\) GeV 
              for \(e^+e^- \to \mu^+e^- +2\tilde\chi_1^0\) in scenario B (circles)
              and  \(e^+e^- \to \tau^+\mu^- +  2\tilde\chi_1^0\) in scenario B (triangles) 
              and I (open circles) for the case of very light neutrinos.}
     \label{fig:ep800}
\end{figure}

\clearpage
\begin{figure}[t]
\centering
\includegraphics[clip]{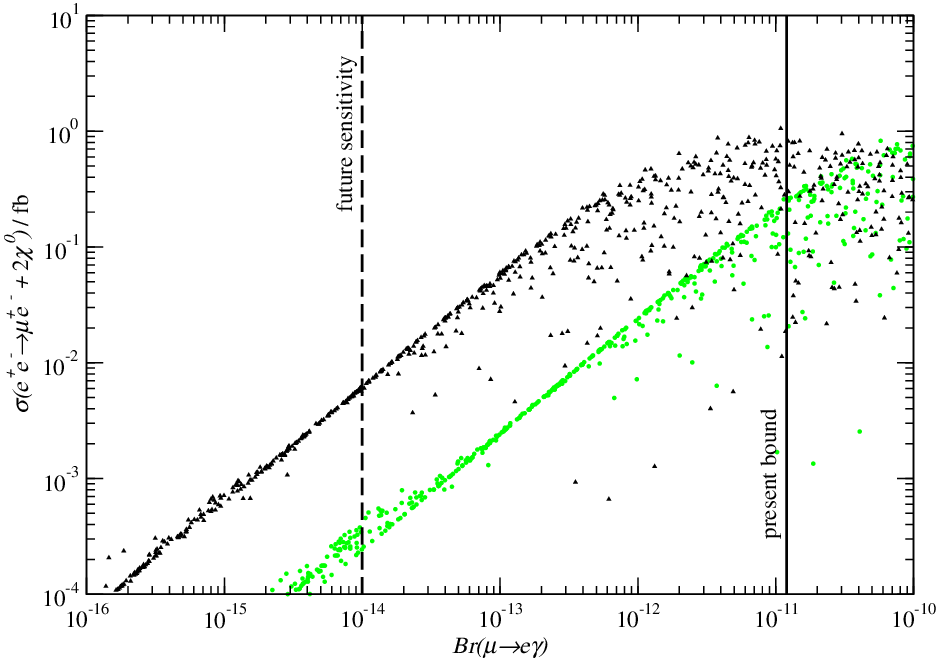}
     \caption{Correlation of \(\sigma(e^+e^- \to \mu^+e^- +2\tilde\chi_1^0)\)
              at \(\sqrt{s}=500\) GeV 
              with  \(Br(\mu\to e\gamma)\) in scenario  C (triangles) 
              and B (circles) for the case of very light neutrinos.}
     \label{fig:emu_lowhigh}
\end{figure}

\clearpage
\begin{figure}[t]
\centering
\includegraphics[clip]{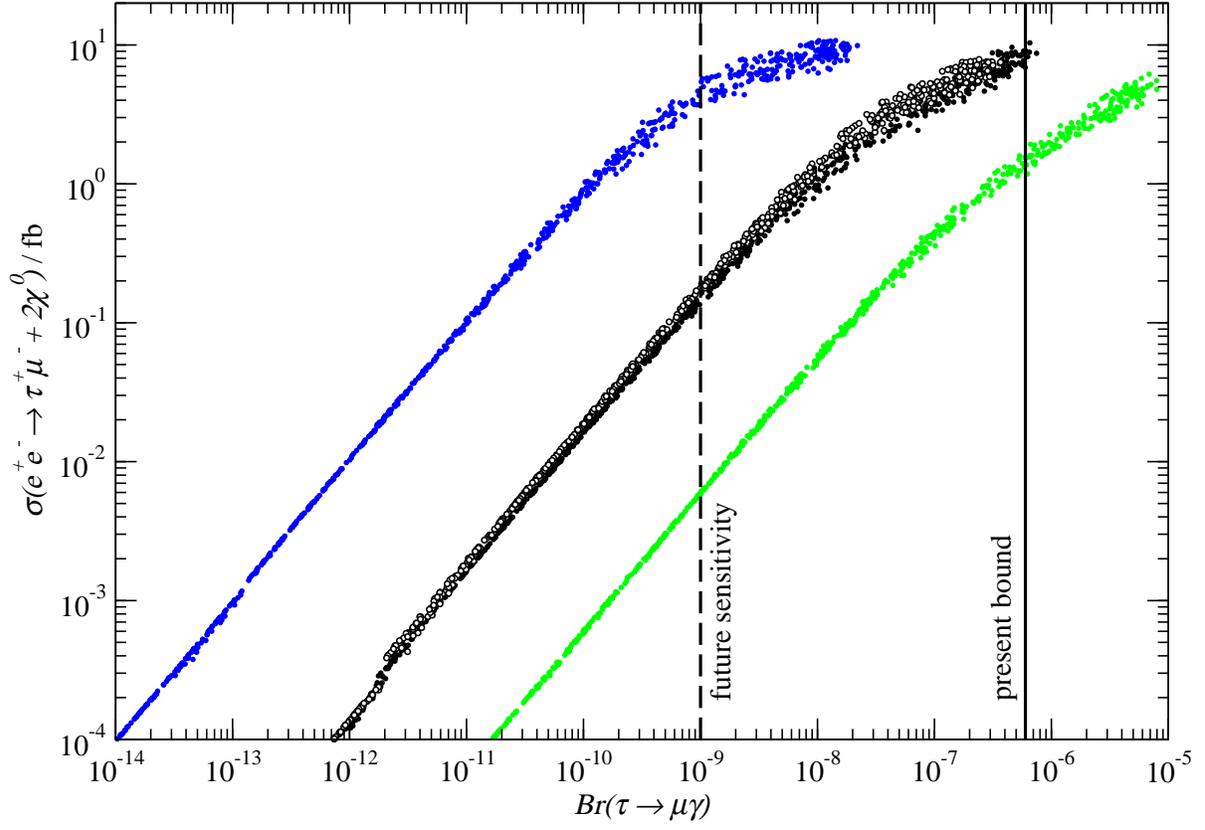}
     \caption{Correlation of \(\sigma(e^+e^- \to \tau^+\mu^- +2\tilde\chi_1^0)\)
              at \(\sqrt{s}=800\) GeV  
              with \(Br(\tau\to \mu\gamma)\) in scenario (from left to right) 
              C, G (open circles), B and I for the case of very light neutrinos.}
     \label{fig:mutau_lowhigh}
\end{figure}
\clearpage
\begin{figure}[t]
\centering
\includegraphics[clip]{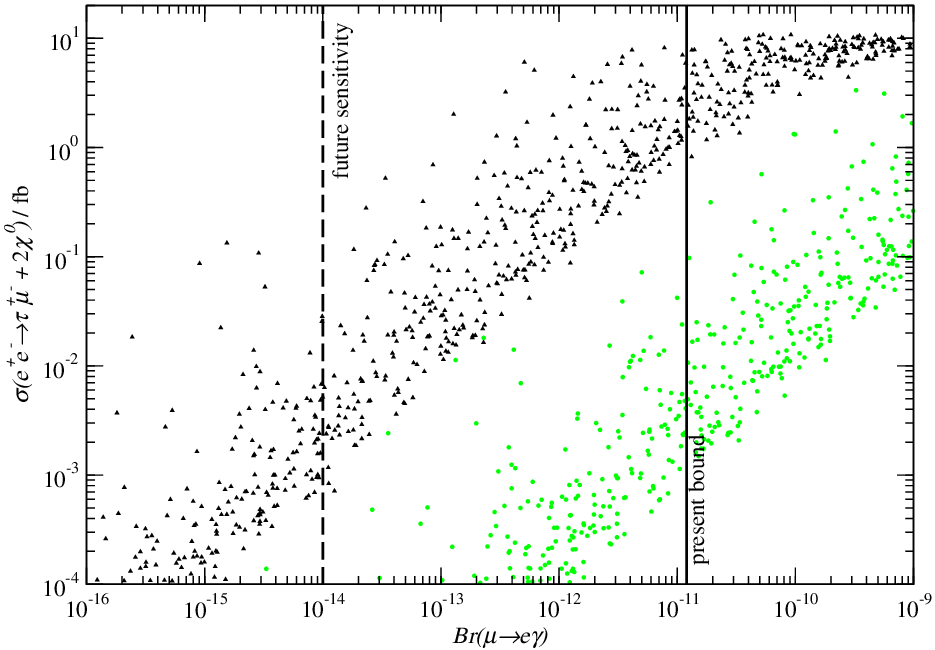}
     \caption{Correlation of \(\sigma(e^+e^- \to \tau^+\mu^- +2\tilde\chi_1^0)\)
              at \(\sqrt{s}=800\) GeV 
              with \(Br(\mu\to e\gamma)\) in scenario C (triangles) and I (circles)
              for the case of very light neutrinos.}
     \label{fig:mutau_emu_lowhigh}
\end{figure}
\clearpage
\begin{figure}[t]
\centering
\includegraphics[clip]{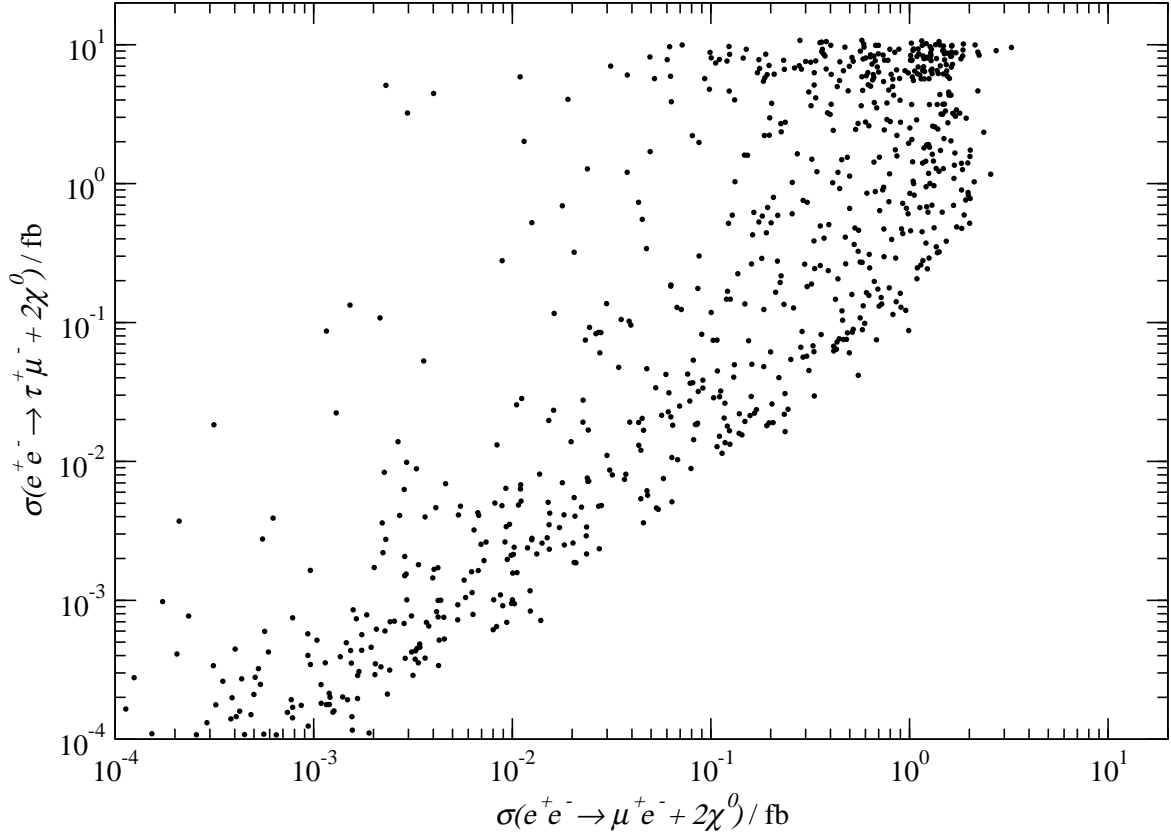}
     \caption{Correlation of \(\sigma(e^+e^- \to \tau^+\mu^- +2\tilde\chi_1^0)\) 
              with \(\sigma(e^+e^- \to \mu^+ e^- +2\tilde\chi_1^0)\) 
              at \(\sqrt{s}=800\) GeV  in scenario C
              for the case of very light neutrinos.}
     \label{fig:highhigh800}
\end{figure}
\clearpage
\begin{figure}[t]
\centering
\includegraphics[clip]{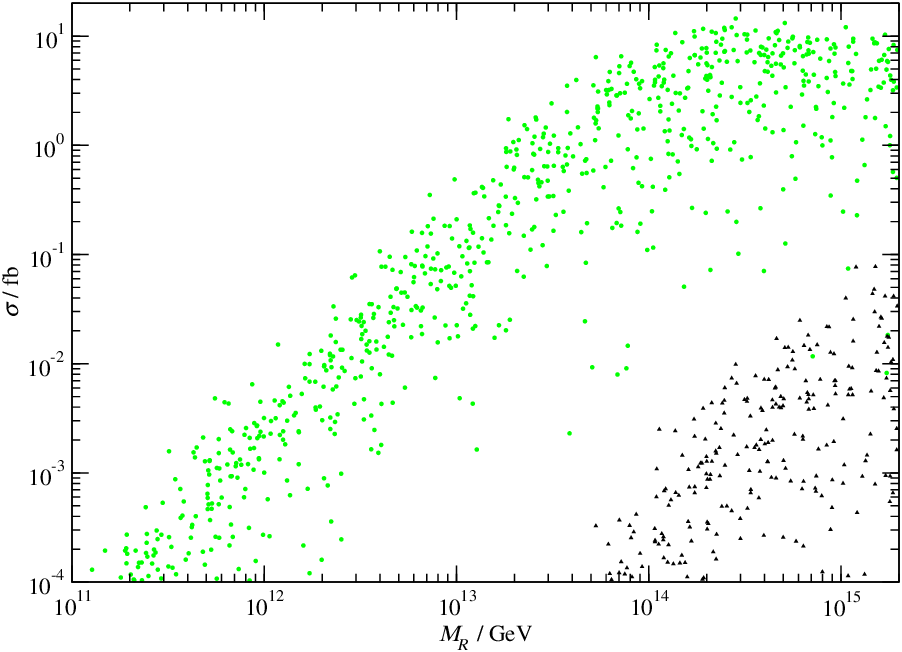}
     \caption{Cross-sections at \(\sqrt{s}=800\) GeV 
              for \(e^-e^- \to \mu^- e^- +2\tilde\chi_1^0\) (circles)
              and \(e^-e^- \to \tau^- \mu^- +  2\tilde\chi_1^0\) (triangles) 
              in scenario C for the case of very light neutrinos.}
     \label{fig:ee800}
\end{figure}
\clearpage
\begin{figure}[t]
\centering
\includegraphics[clip]{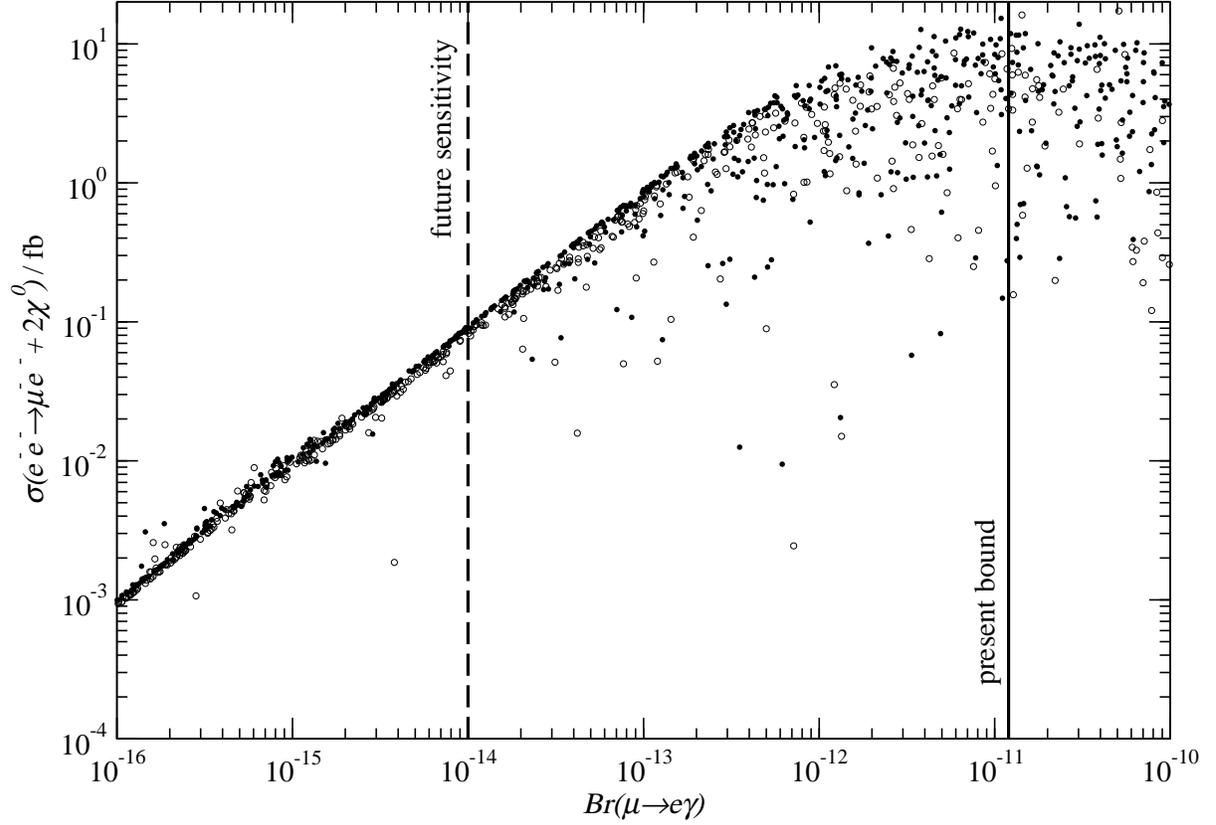}
     \caption{Correlation of \(\sigma(e^-e^- \to \mu^- e^- +2\tilde\chi_1^0)\)
              at \(\sqrt{s}=800\) GeV with \(Br(\mu\to e\gamma)\) 
              in scenario C for the case of very light neutrinos (closed circles) 
              and degenerate heavier neutrinos (open circles).}
     \label{fig:emu_lowhigh_ee}
\end{figure}

\end{document}